# Biological Effects of Microwaves: Thermal and Nonthermal Mechanisms*

A Report by an Independent Investigator

**by John Michael Williams**

`jmmwill@comcast.net`

**Wilsonville, Oregon**

**2016-09-28**

\* Some of the present paper was submitted in 2001 to the U. S. Federal Communications Commission as a *Comment* on unlicensed operation of ultrawide-band devices: `http://arXiv.org/pdf/physics/0102007`. Also available for download at `http://www.scribd.com/jmmwill/documents`.

# Abstract

For over sixty years, it has been reported that microwave electromagnetic radiation (EMR) had effects on humans which could not be explained by detectible heating of tissue. Auditory responses to radar, called *microwave hearing*, have been the best known of these phenomena. To account for microwave hearing, many studies in the literature have adopted a rate-of-heating hypothesis advanced by Foster and Finch in 1974. We show here that theoretical and experimental studies supporting this hypothesis are weaker than usually assumed.

We develop a simple framework of understanding of EMR that may be used to explain microwave hearing as a nonthermal, nonacoustic effect. We then extend this approach to other contexts, pointing out several fundamental misconceptions confounding the field.

EMR, especially wide-band EMR, primarily must have a nonthermal effect on living tissue before conversion to heat. Auditory and tactile sensations, central neurological disability, and blood pressure loss caused by EMR have been documented. Except microwave hearing, parameters of irradiation causing such effects have not been explored adequately and remain unknown. There appears not to be any forensic methodology to prove the cause of harm at nonthermal levels.

# Table of Contents

(This page intentionally blank)

# Introduction

*A good theory should be as simple as possible.   But no more so.*

-- Albert Einstein

## *Nontechnical Controversy*

Claims that electromagnetic radiation (EMR) from radar or radio communication devices might have subtle biological effects have in the past opened the door to speculation and to ignorant as well as unwarranted accusations.   The problem is that EMR is invisible and so can not be shown absent when a putative victim has experienced some personal harm -- for example because of crashing an auto or a plane, forgetting an appointment, or experiencing otherwise unexplained pain or disease.

Radar operators in the late 1940's sometimes remarked that they could hear things such as clicks or buzzes when standing in or near the beam of a transmitter (Chou *et al*, 1982).   Because safe radio-frequency irradiation dosage levels were not well understood, experimentation was strongly discouraged.   At that time, and perhaps even now, there was folklore that radar could have no effect but destructive heating, and that such sensations must therefore be an indication of mental incompetence.

Since the 1940's, there have been isolated and vague, often apparently paranoid, complaints of "mind control" or Nazi-like experiments using secret drugs or hidden electronic apparatus.   U. S. Congressional inquiries (Ervin, 1974; Rockefeller, 1994) substantiated some of them and mandated new patient safeguards in medical research.   Thus, the history suggests current or future similar claims about microwaves.

As documented in many peer-reviewed studies below, and experienced by the present author, microwaves can be heard.   The apparent loudness can be almost deafening.   Microwaves also can have distinct, often painful, tactile effects not resembling anything attributable to heat.   Some of the author's own experience in measuring these effects has been posted online (Williams, 2013).   A correspondent of the author claims he has observed handheld radar sets being used to flush deer from underbrush during hunting season.   As will be shown, whereas such claims individually might be true or false, there is no more reason to doubt them *a priori* than to doubt the sanity of radar operators.

More recently, concerns have arisen about the safety of cell phones (Stewart, 2000 as updated by Swerdlow, 2003; Hyland, 2001; Adey, 2003; see also Barnett, 1994), and even of LORAN transmissions (Dawes, 2001).   These concerns, actually often fears, are being stoked by rumors about covert use of portable radar by soldiers or police (Burton and Ohlke, 2000; Eisley, 2001; Jones, 2005), and by public but ambiguous announcements of military weaponry based on microwaves (Sirak, 2001), although such weaponry, even if effective, likely would be illegal in military engagement (Williams, 2001).

The underlying fear here revolves around the possibility of serious intentional injury or death delivered by microwave transmitting devices, with no forensic methodology to recognize the symptoms or identify the cause.

There does not seem to be any possible political resolution of these controversies, which are based often on empty suspicion about unobservable media; the only rational solution is greater understanding of the claims, based on a good understanding of the biological effects of EMR.

To lay groundwork for understanding, we focus on well-established nonthermal biological effects of microwaves, explaining how a thermal interaction might occur; then, we criticize the spread of heat (not light) which has replaced good physics in guiding research in this area.  Throughout, we apply a rationale which should lead readers out of the morass of confusion into which many otherwise rational solutions have been collected.

## *Technical Controversy*

Very little about the biological effects of EMR on living tissue appears in the literature until Frey (1961; 1962) announced that he had discovered some peculiar tactile and auditory effects while exposing humans to pulsed radar at average power far too low to heat tissue significantly.  The auditory effects, including clicks, hisses, or buzzes, were easily demonstrated; they were dubbed *microwave hearing*.

Interest in the engineering and biomedical communities increased gradually, with numerous theoretical interpretations, and experiments on humans and animals, appearing through the late 1970's.  A very thorough categorized list of the literature was published by Glaser (1972).  We shall not review the literature here, but we refer the reader to the wide range of experiments published in a special supplement of *Radio Science* in 1977, and to reviews by Frey (1971), Adey (1981), Chou, *et al* (1982), Erwin (1988), Lai (1994) and Foster (2000).  Related government studies have been published in Barnett (1994), U. S. Office of Technology Assessment (1989), Stewart (2000), Hyland (2001), Swerdlow (2003), and World Health Organization (2007).  The problem also is of concern in Russia (Kositsky, *et al*, 2001).

Several hypotheses concerning microwave hearing were advanced to support one idea or another.  Almost all the evidence indicated that bulk heating could not be responsible for the easily demonstrated tinnitus-like auditory effects of pulsed EMR.  The issue apparently was resolved by Foster and Finch (1974), hereafter abbreviated FF74, who concluded that *rate of change of temperature* of tissue water could convert an EMR pulse to an acoustic one of very small amplitude, but large enough effectively to stimulate the cochlea.  Thus, all biological effects of EMR either would be by temperature increase (at high average power levels) or by rate of temperature increase (at lower average levels).  At high levels, the effects were predictable and harmful; at low levels, the effects were not predictably harmful.

Papers by Frey and others argued against the FF74 hypothesis.  Thermal and other mechanisms were discussed by all concerned on several forums, notably as reported in Guy, *et al* (1975).  The controversy reached its height in *Science* magazine (Frey and others, 1979); but, over time, the majority of researchers came to accept the FF74 hypothesis, although Foster himself maintained some reservation about its exclusiveness (Foster, 2000).

# Some Biophysical Perspective

Explanatory note: The following biophysical approach is believed correct and complete insofar as it relates to the matters being discussed in this paper. An understanding of the biophysics deepens the understanding of the EMR mechanisms at work; however, this understanding is not itself necessary to comprehend the arguments below which criticize the elementary physical errors rampant in the EMR literature.

Readers uninterested in these details may skip to the "Problems With Thermal Causation" section below.

## *Membrane Functionality*

The body cells, especially the individual nerves and sensory receptors, function as ordered systems of boundaries in fluid. The mechanism of action of nerves, muscles, and systems of them, has been described in terms of membrane polarization changes (Regan, 1971, ch. 1). These polarization changes occur because membranes not only are populated by structure-related dipoles as in the figures below, but also are more sparsely populated by specialized channels which can open or close to regulate an exchange of ions. These ions primarily are sodium (Na+) and potassium (K+), but also may include calcium or others.

Living cells internally generally are more negative than their local environment. They thus store some free energy in the electric potential across their bounding membranes. This potential difference is electrochemical and depends on continuous expenditure of energy to maintain higher intracellular K+ concentrations and lower intracellular Na+ concentrations than would be so during chemical (osmotic) equilibrium. The ion gradients are controlled by enzyme action which does its work using the adenosine triphosphate (ATP) energy store available to cells throughout the body. At rest, the membrane potential is maximal and therefore must maximally align dipoles in the membrane, so that the cumulative dipole moment orthogonal to the membrane surface also is maximal.

Specifically, resting neurons expend energy to operate a Na-K "pump" which keeps the potential inside the cell about 70 mV more negative than the outside (Regan, 1972, p. 7). An additional osmotic potential also exists. The membrane can liberate energy very locally by allowing some extracellular Na+ to flow in and some intracellular K+ to flow out. Dendritic depolarization (*e. g.*, "C" nerve fiber sensory response) may cause the –70 mV potential to drop close to 0 mV immediately adjacent to the depolarized areas of the cell membrane.

When the originally resting electric field across a nerve cell membrane develops a steep enough spatial gradient parallel to the membrane surface, some closed channels are opened, causing a spread of depolarization which may become an action potential under proper circumstances. The sudden depolarization during an axonal action potential may cause the resting potential of –70 mV to swing past 0 perhaps to +10 mV. Local hyperpolarization may reduce, modulate, or prevent the generation of regular action potentials which maintain the state of the central nervous system. The equation of locally <u>steep</u> spatial gradients with locally <u>sudden</u> temporal ones is implied by diffusion and is essential in the Hodgkin-Huxley and similar approaches: For ion species *j* in concentration $u_j(x,t)$,

$$D\nabla^2 u(x,t) - \rho u(x,t) = \frac{\partial u(x,t)}{\partial t} \text{; so, over time, } \nabla^2 u - (\text{loss}) \propto \frac{\partial u}{\partial t} - (\text{system input}), \quad (1)$$

a linear description, but one adequate to show the principle. Eccles (1973) and references therein may be consulted for details on initiation and propagation of neural action potentials. We shall relate membrane spatial gradients to EMR action after describing the effects of EMR on membrane polarization and discussing their implications.

## *EMR and the Human Body*

Unless otherwise noted, the EMR under discussion should be assumed to be in the microwave frequency range. Because coupling of magnetic fields to tissue is far weaker than that of electric (*E*) fields, we shall not discuss the magnetic fields which always accompany the *E* fields of EMR. Works such as those of Binhi and Chernavskii (2004) may be referenced on such topics. Here, we merely remark that a changing magnetic field can induce an *E* field which may cause thermal or nonthermal biological effects of interest to us.

EMR is attenuated exponentially with depth by any uniform body tissue. For example, a reasonably accurate equation relating the field intensity at depth *d* to the applied intensity in air, might take the form,

$$I_{\text{tissue}}(d,f) = I_{\text{air}} \cdot \exp(-k \cdot d \cdot f^{1/2}), \quad (2)$$

for *k* some tissue- and unit-dependent constant, and *f* the frequency. Penetration increases with decreasing frequency.

A zero frequency, direct-current (DC) voltage applied across the body will cause current to flow by ionic conduction in all conductive body fluids. When the DC coupling is through air -- for example by placing the body between two conductors at different voltages, but not touching -- the current flow quickly will stop because of the resistivity of air; but, displaced charges in the body, and in the coupling electrodes, will remain displaced to cancel the fields applied to those charges. The same occurs with the body replaced by a piece of conductive metal, such as aluminum.

When microwave EMR is applied to a piece of metal, most of it will be reflected; and, so, the penetration will be small. For such a medium, the internal field caused even by an applied DC potential will be close to zero. However, the reason the penetration in a metal is small is that the charge carriers in the metal, electrons, respond so actively that they cancel the field by their motion (in the case of EMR) or by their static relocation (in the case of DC).

For the moderately conductive human body, a large fraction of which consists of conductive saline water, the penetration of EMR is limited by absorption as well as reflection.  One expects from Eq. (2) that any biological effect, an oscillating flow of ions, will be large near the surface and will decline about exponentially with depth.

Regardless of penetration depth, we shall suggest in this work that whenever EMR is applied to well-ordered tissue or organ boundaries of the body, well-ordered interactions must be taking place before the end-result of heating takes place.  The difficult question is <u>how to apply the EMR</u> so it is amplified by the living organism to a level unequivocally observable in a scientific experiment.

## Dipole Rotation by EMR

A microwave EMR beam may be viewed as imposing time-varying forces on charged elements, monopoles, in tissue by virtue of its oscillating $E$ field.  Such charged elements may include atoms (ions), molecules, regions of bonding between molecules, or anything else with a localized excess of positive or negative charge.  Such localized excesses of charge always are cancelled by oppositely charged elements not far away; tissue never has a significant, long-range net excess charge when compared with the total number of cancelled-out, neutralized, elements of which it is composed.  On the scale of monopole spacing distances in tissue, a microwave beam presents a time-changing but spatially uniform $E$ field to those elements; the vector potential also is effectively uniform on this scale, so tissue will have no direct magnetic interaction at all with the EMR (*e. g.*, Craig & Thirunamachandran, 1984, Sections 3.6 & 4.7).

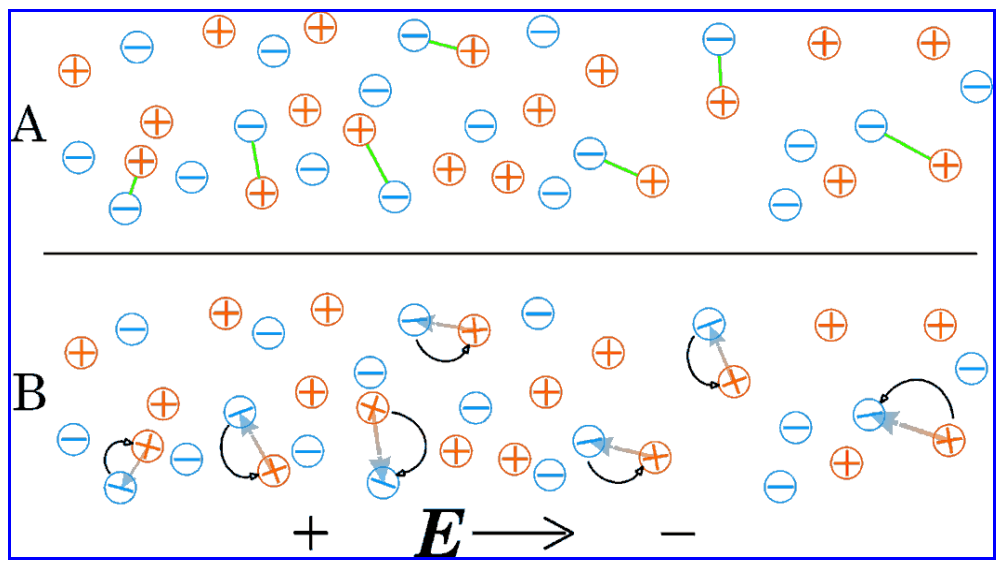

**Figure 1.  A:  Charged elements in a matrix, some bound (straight lines) into dipoles.  B:  The effect of an E field in terms of rotating dipoles (straight arrows represent dipole moments; curved arrows represent rotation).**

In whatever matrix in which they are found, fluid or not, monopoles of opposite charge may be connected more rigidly together as a *dipole* than to other nearby elements. Dipoles typically are distinct molecules, but not all molecules are dipoles.

If a uniform *E* field is applied parallel to the line between the poles in a dipole, it will cause attraction of one pole and exactly equal repulsion of the other, causing some insignificant bond length change coupled with a zero net displacement of the dipole in the matrix.   If the field is applied in a direction perpendicular to the line between the two poles, the poles will be displaced in opposite directions and the dipole then will be rotated.   In this case, the distance between the poles will not have to change, so no work will have to be done by the EMR field in the relatively strong potential binding the poles. Furthermore, the dipole itself, being neutral, will not be displaced in the matrix, requiring no work done there, either.   This is the concept of dipole rotation by EMR illustrated in Fig. 1 above.

In empirical fact as well as theory, weak EMR of relatively long wavelength locally affects nonmetallic objects just as described:  It rotates dipoles if they have time to respond to the field oscillations, but it stretches bonds between monopoles insignificantly. If the oscillation frequency is adequately low, the net effect continually is to change alignment of the dipoles against the field so that the EMR field locally always is kept minimized in the tissue, given that the EMR force is cancelled.   See the section below on "Time Domain Patterns" for a quantification of this approach in terms of stimulation of human neural response.

## Conversion of EMR to Heat

Consider a tissue boundary such as the mucous membrane of the stomach wall or the endothelium of a large blood vessel.   Such a boundary will be enclosed in electrically polarized cell membranes with a regular coulomb structure, the orientation of the membrane dipoles being completely organ dependent in the absence of EMR.   How could EMR raise the temperature of such a structure?

Suppose we had a coherent microwave beam producing a standing *E*-field wave, as in a microwave oven:  How would we understand the effect of the EMR?   We shall disregard here the phase-mixing technology in a real microwave oven as infinitely slow when compared with its 2.4 GHz cycle rate.   We know that an EMR standing wave pattern can diffract moving, charged particles into highly organized patterns (Nairz *et al*, 2001), so why should a standing pattern cause random motion -- heat -- in anything?

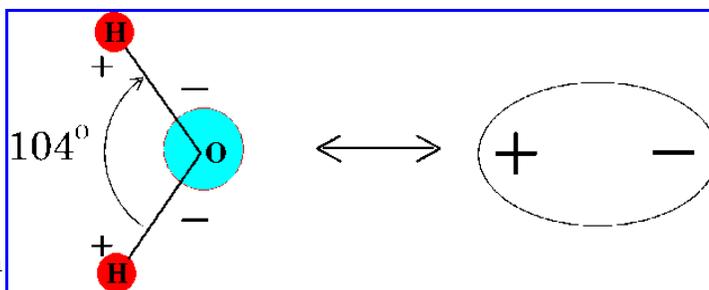

**Figure 2.  Simple, idealized water molecule. Left:  The planar arrangement of atoms in a water molecule.   Right:  The equivalent electric dipole element.**

First, let us examine the action on the fluid enclosed by some such tissue boundary.   How would microwave EMR raise the temperature of the fluid?

As shown in Fig. 2, a simplified explanation invokes the structure of the water molecule, which consists of an oxygen atom covalently bonded to two hydrogen atoms. The hydrogens interact and are drawn together somewhat, so that they are almost at right angles. The hydrogens are positively charged, and the oxygen negatively, so a triangular, planar arrangement equivalent to a dipole results.

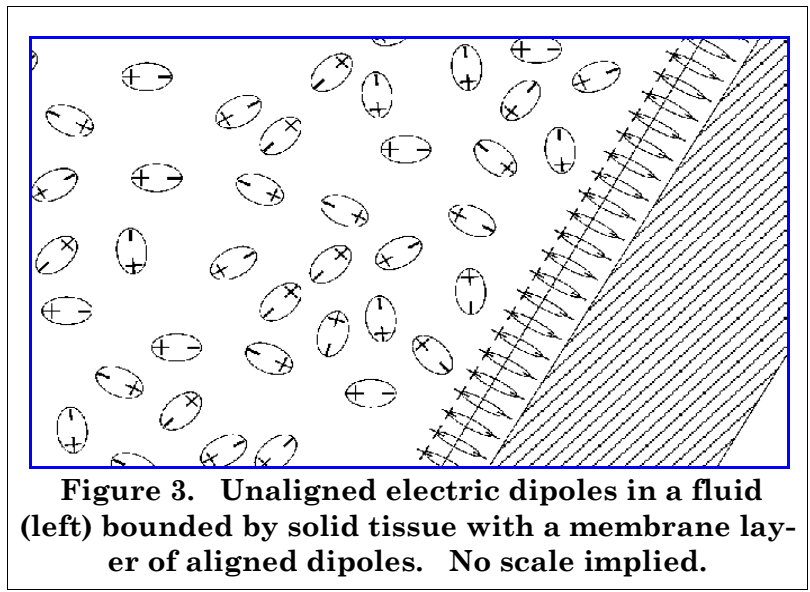

**Figure 3.** Unaligned electric dipoles in a fluid (left) bounded by solid tissue with a membrane layer of aligned dipoles. No scale implied.

Some discussion of the structure of an isolated water molecule may be found in Tsiper (2004). An aqueous fluid then may be viewed as a collection of electric dipoles. For example, without much loss of generality, let us take a structure bounded by a solid tissue with its electric dipoles aligned with positive poles locally pointing into the fluid. This is shown in Fig. 3: Let each fluid dipole in Fig. 3 represent a water molecule, as modelled simply above. The dipoles initially may be assumed arranged more or less at random with respect to one another, and in random thermal relative motion. When they are rotated (torqued) together, all magnitudes in phase with the EMR, the dipoles interact, as expected, at random distances and phases with respect to one another.

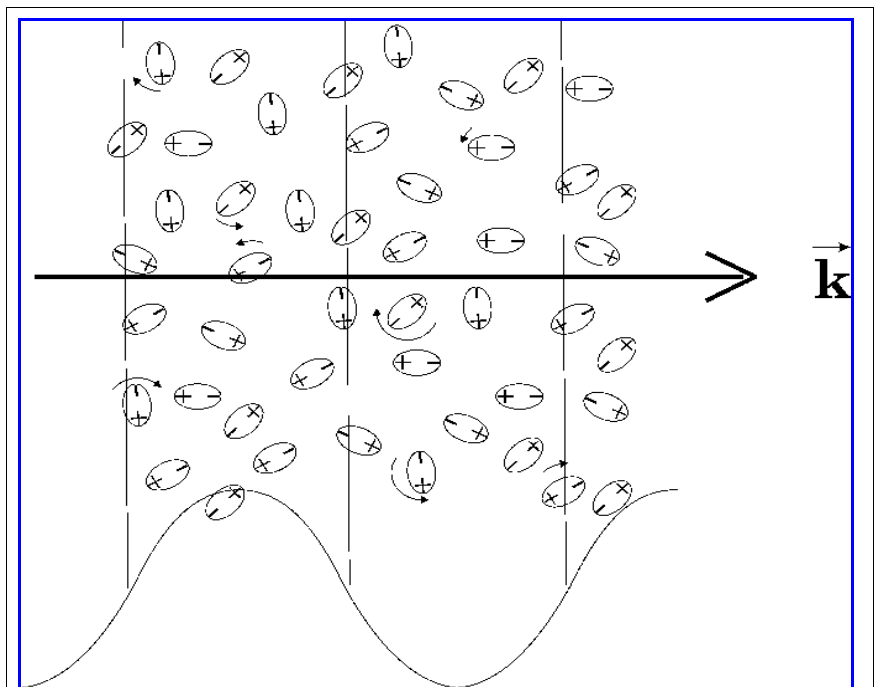

**Figure 4.** Sketch of the effect of a plane EMR wave propagating in a bulk fluid of polar elements, with a net result of increasing the temperature. The *E*-vector magnitude and polarity (+ up) is shown on an arbitrary vertical scale drawn through nodes of the wave; the horizontal scale is for illustration only. The EMR torque on a few of the polar elements is shown as a vector in the $\pm\vec{H}$ direction.

For this reason, then, even coherent EMR-induced motions must contribute random momentum and therefore random energy. This is shown in Fig. 4, where the wave vector $\vec{k}$ is drawn for EMR propagating as a plane wave from left to right, with a few illustrative EMR torque angles shown for individual dipoles. The wave vector is given only to define the direction of the incident EMR as that of $\vec{E} \times \vec{H}$ in our simplified representation.

The EMR energy converted by a polar fluid, then, is completely disordered on the scale of anything macroscopic which we might insert into the fluid; therefore, the EMR energy is heat energy.

## Conversion of EMR by the Body

The human body does not consist of a bag of fluid but rather of a system of organs, each with its own, well-defined boundary from the others. The different tissues in different organs each have their own dielectric properties, as described in Durney, *et al* (1999). The accurate representation of human response to microwave radiation is exceedingly complex, as discussed and approximated in Cundin and Roach (2010). In any case, organ boundaries may function as diffractive, refractive, and reflective interfaces for microwave EMR of wavelength comparable to the boundary extent.

On a finer but still macroscopic scale, let us next consider EMR near the fluid boundary on the right of Fig. 3 above. The boundary is more ordered than the bulk of the fluid. The effect of EMR is sketched in Fig. 5 below, where a vertically limiting beam envelope is assumed. The amplitude of the beam in that figure is maximum near the horizontal line near the middle of the figure on the left; this amplitude dies off away from the middle, upward and downward, perhaps as a Gaussian function.

Near the fluid boundary in Fig. 5, some EMR energy will be delivered to the membrane as mechanical energy, doing work on the membrane associated with the linear and angular momentum of the EMR, which therefore would not be work associated with heat dissipation.

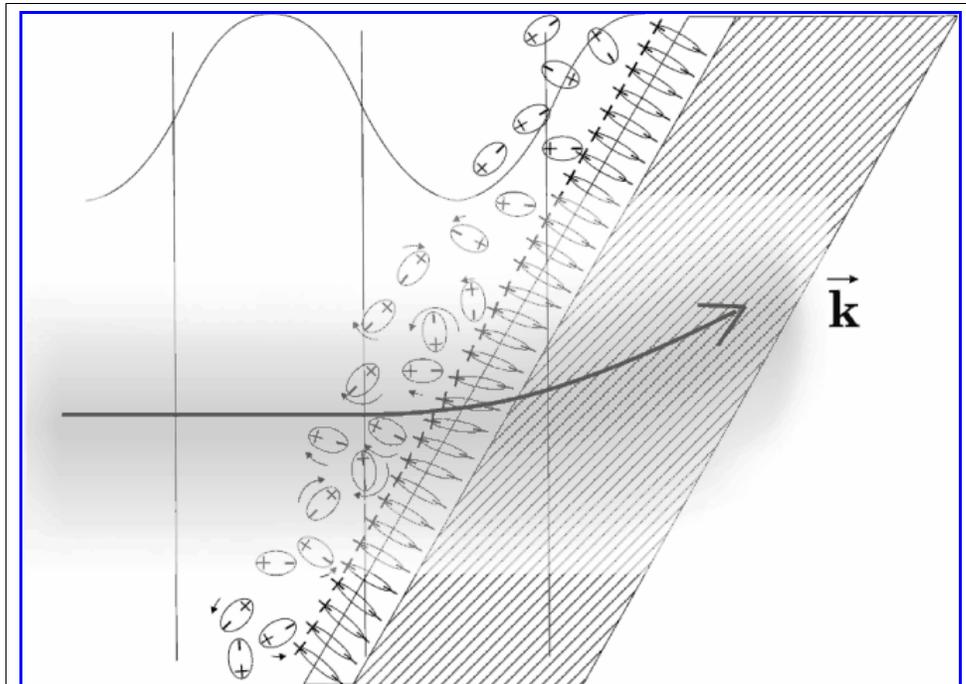

**Figure 5.** Sketch of the effect of an EMR beam propagating from a polar fluid into tissue bounded by a polarized membrane. Nothing to scale. The refraction proves that some work will be done on the boundary, so that some nonthermal energy will be transferred.

Otherwise stated, this mechanical work would be equivalent to free energy delivered by a heat engine, not to heat merely transferred by a temperature difference according to the laws of thermodynamics.  It is not physically possible that no free energy would be delivered, because delivery of pure thermal energy would require complete disorder in the spacing and orientation of the bounding polar elements.

So, at least some energy transferred to the body by EMR will be not be thermal, because of body structure.  The EMR will impose a pattern with spatial periodicity approximately equal to half of the projected wavelength (in the surrounding medium) on all boundaries.  Polarized molecules in the bounding membrane but away from the EMR nodes will be rotated on temporal average away from their rest positions.  Because of reduced cancellation of the ionic potential, the time-averaged membrane polarization on the boundary thus will be increased except at nodes of the EMR.  For nerve cells or some muscle cells, the membrane will develop a changing spatial hyperpolarization pattern possibly interfering with the normal changes responsible for its normal functionality.  We shall discuss this last in more detail later.  Ingalls (1967), for example, adds in a final note that the same pulsed radar which humans could hear also reduced the measured nerve impulse threshold in the dissected sciatic nerve of a frog.

If we assume in general that the wave vector $\vec{k}$ will not be perfectly perpendicular to the boundary, then the interaction in Fig. 5 (above) will deliver some linear momentum transverse to $\vec{k}$; the EMR beam will be refracted.  The angle of refraction will be related to the amount of free energy delivered to the boundary dipoles.  The right side of the figure shows this refraction as a bending upward of the wave vector.

The momentum transferred by the reaction force of the refracted beam will be delivered to the membrane and will correspond entirely to free energy, doing work by energy associated with linear and angular momentum from the microwave photons; no heat will be involved.  Other energy may be transferred by the EMR, depending on the nature of the medium at and beyond the boundary.  Absorbed energy need not be thermal; in terms of broad categories, it might be mechanical, optical, chemical, biological, *etc*.  We ignore here the dielectric and other properties quantifying the direction and amount of refraction.  We also merely mention diffraction and reflection(s), which might set up standing wave modulations in the incident beam.

From all this discussion, it seems clear that the burden of proof should be on any claim that microwaves act like microwave ovens and do nothing but cause heat to be transferred to living tissue.  Because human tissue is formed into organs with distinct boundaries, based on the *prima facie* argument of this section, EMR should be assumed to have nonthermal effects on living tissue unless explicitly provable otherwise.

## Complexity of Water and Cellular Contents

In this section, we lean heavily on Chaplin (2002) for a better understanding of water. Individual water molecules maintain their identity at ordinary temperatures for perhaps a millisecond (Chaplin, 2002), but pairs of them are intrinsically unstable.  There is no way two planar water-molecule dipoles could line up in the three dimensional bulk of a fluid so as to have the negative (oxygen) pole of one stably associated with the positive (dual hydrogen) pole of another; therefore, larger groups of water molecules have to be formed.  In actuality, the hydrogen atoms in each molecule form hydrogen bonds which chain these groups of molecules together in all directions.  Because water is liquid at biological temperatures, groups of molecules are formed only transitorily; but, as discussed in detail by Chaplin, large assemblages can persist on the average for relatively long times.

The theory of water by no means is settled, but Chaplin presents a convincing and widely accepted way of explaining many of the puzzles of water:  The explanation postulates that the dominant structure in liquid water is a spherically symmetrical one, an icosahedron of twenty 14-molecule tetrahedra, with volume containing a total of 280 water molecules.  The tetrahedra are described in Krisch, *et al* (2002); dodecahedra apparently are resolved by Miyazaki, *et al* (2004) and Shin, *et al* (2004).  In water, regardless of the precise structure, the time for a hydrogen bond to be created is about $10^{-4}$ ns (Asbury, *et al*, 2003).  Individual hydrogen bonds persist without breaking apart for perhaps $10^{-2}$ ns and usually re-form in the same relation in which they were formed before breaking.

The icosahedra are a huge 3 nm in diameter, about 1/3 the thickness of a typical cell membrane.  Very approximately, a compressional sound vibration at 1500 $\text{ms}^{-1}$ in water would propagate across the diameter of such an icosahedron in about $2 \cdot 10^{-3}$ ns, far shorter a time than the lifetime of one of the hydrogen bonds keeping the icosahedron intact.  Other vibratory modes of sound would propagate more slowly than the compressional mode, but it seems clear that such icosahedra could sustain internal vibrations which greatly would reduce the heat energy made available immediately from an absorbed microwave photon.

Larger structures of 13 icosahedra are believed to form.  Smaller, transient linear chains form, and so do fragments of these latter; but, the icosahedra tend to persist longest because they are the minimal structures with complete stability conferred by their symmetry.  A simulation study of a dipole model of liquid water by Higo, *et al* (2001) confirms that water should contain stable structures approximately corresponding in size to the postulated icosahedra.  The Higo, *et al* simulation was run only for 1/3 ns, but the result suggests that these structures should be expected to persist for some indefinite time greater than 1 ns.  Other simulations not producing such long persistences have been performed but typically are based on assumptions of spherical symmetry *a priori* expected to negate the correlations being sought (*e. g.*, Kumar, *et al*, 2005).  Theoretical flaws implied by spherical symmetry are discussed in a different context below.

Returning to the Chaplin model, each icosahedron has two main conformations, an expanded state (ES) and a collapsed state (CS). The ES is a true icosahedron, everywhere convex, but the CS has one or more vertices buckled inward, creating small local concavities. Chaplin presents figures clearly showing this difference. The equilibrium between ES and CS depends on temperature. At lower temperatures, ES dominates; at higher temperatures, CS dominates. The volume occupied by 280 water molecules is less in the CS state than in the ES, so the effective density of CS is greater than that of ES. To explain the empirical 4 C density maximum of water, one merely assumes that density increases between 0 C to 4 C because some ES moves over to CS. Above 4 C, yet more CS occurs, but fragmentation increases also; every stable structure possesses independent thermodynamic degrees of freedom. Because of fragmentation, the density therefore decreases above 4 C.

Because most of the content of a body cell is $H_2O$, the cellular contents must be assumed built upon a structured water substrate, in addition to being highly structured themselves.

### EMR Effect on Water

Application of EMR would be expected to affect molecular dipoles in liquid water much the way it affects dipoles on a tissue boundary: If below some limiting power level, and in a low enough frequency range, EMR should rotate dipoles in the icosahedra or other longer-range structural units, bending hydrogen bonds and deforming them, but should not destroy those units. A cyclic but constrained rotation of dipoles should cause elastic deformation waves within any large-scale water structure.

Under such conditions, then, EMR only should be converted to heat at the random boundaries among the water structural units, and this should occur only in a relatively small fraction of the volume of any given quantity of water. The main fraction of the EMR energy input would persist in bond deformations or internal vibrations. It should be easy to see that in the limit of a completely structured object, such as a single, solid piezoelectric crystal, a nondestructive pulse of EMR merely would deform the lattice elastically, causing no immediate temperature rise at all.

Thus, conversion of EMR to heat in water under conditions of limited power would be expected to develop relatively slowly, as the rotated structural dipoles propagated mechanical changes from the water structures to their structurally random boundaries, and as the limited lifetimes of the water structures caused them to break up. Consideration of the structure of water will be applied in the discussion of the FF74 experiment below.

## Interaction with Macroscopic Body Geometry

EMR interacts not only on tissue boundaries but also by antenna-like interaction with the gross dimensions of the human organs. The "resonance region" (Foster, 2000) for the human body varies with the organ or body part involved. In this region, between frequencies of about 30 MHz and 500 GHz, wave lengths and body part lengths may cause complicated near-field interactions difficult or impossible to calculate.   We shall refer to this as the *diffractive* region of the spectrum, keeping in mind that, in it, the local geometry of body structures can concentrate or attenuate fields -- perhaps by as much as a factor of ten.   We shall not distinguish interference from diffraction in this usage.

An antenna-like amplification of EMR by the human body can be demonstrated easily with a simple monopole antenna and an instrument such as an oscilloscope: Attach the antenna and observe the normal, ubiquitous background RF noise as displayed by the instrument.   Then, grasp the antenna with the hand:  The amplitudes will increase by a factor of two to ten, depending on conditions.   The three-dimensional human body is far better a generic antenna than any one-dimensional monopole of comparable length.

In the human body, there will be acoustic resonances, too.   Given a sound speed of under 600 m/s in the normal human head (Bekesy, 1948), a 5 kHz sound wave, well within the normal human auditory range for sensation of a hiss or crackle, would have a wavelength of about 12 cm.   This would be in the *acoustic* diffractive range for a region about the size of the human head.

So, at frequencies around 1 GHz, there would be both an electromagnetic and a possibly fed-back acoustic contribution to EMR diffractive influence in the human head. This coincidence would not hold the same way for animals with substantially smaller heads, such as the majority of dogs, or for rhesus monkeys, cats, rats, or mice.   Sound waves in the audible range produced in the human head by any mechanism would find the geometry conducive to diffractive effects such as coherent superposition.

It is unknown whether auditory response to microwaves actually would be increased by the combined effect of the EMR and acoustic diffractive effects; to the present author's knowledge, no experiment or simulation has addressed the question.   However, the use of microwave-produced acoustic resonances to be recorded by external diagnostic instrumentation is described in Wang *et al* (1999) and Xu *et al* (2001), the former reporting that the microwave-induced acoustic pressure was proportional to the microwave intensity.

It should be mentioned here that some initial applications of microwaves in mineral analysis and mining are described in (Monti, *et al*, 2015), although human responses were not measured.

# Problems With Thermal Causation

We first discuss the FF74 study, an approach which has provided the best and most elegant result in support of a thermal basis for biological response to microwaves. We suggest that the argument for an exclusively thermal cause of auditory stimulation is weaker than previously has been recognized. We then point out problems with supporting arguments often advanced to confirm the Foster and Finch hypothesis, but which are not central to it.

## *Microwave Hearing: The Foster and Finch Hypothesis*

To produce microwave hearing according to Foster's (2000) description, pulses optimally would be about 5 $\mu s$ wide, in the frequency range 1 to 10 GHz, and should deliver perhaps $\sim 10^4 \, \text{W}/\text{m}^2$; this is the power during the pulse.

Note that $10^4 \, \text{W}/\text{m}^2$ is the same as $1 \, \text{W}/\text{cm}^2$. Each $5 \, \mu s$, $1 \, \text{W}/\text{cm}^2$ pulse produces the sensation of a distinct "click" or "pop" when directed at the head of a human with normal hearing for high frequencies. The loudness, according to Frey (1962) and others, may vary from barely perceptible to 60 dB or more. At 5 pulses per second, the duty cycle would be only $25 \cdot 10^{-6}$, making the time-averaged power just 25 $\mu\text{W}/\text{cm}^2$.

At 1 GHz, the EMR wavelength would be about 30 cm in air; at 10 GHz, 3 cm -- all in the diffractive range for the human head. Under the FF74 hypothesis, in response to an EMR pulse, the head would vibrate as a whole, and the microwave hearing would be an acoustic response mediated by bone conduction to the cochlea. As an aside, we add here that there also might be acoustic diffractive contributions, as suggested above, unique to a human-sized head.

According to FF74, the thermoacoustic effect is supposed to be mediated by the power transferred as a minute temperature rise during each pulse. A 5 $\mu s$ pulse would contain some 5,000 to 50,000 cycles of the carrier at 1 to 10 GHz, respectively. Because, then, according to this theory, almost none of the heard microwave energy could be delivered by the pulse edge transients, the thermoacoustic effect should be described as entirely a pulse envelope (low-pass) phenomenon. This alone would imply a different mechanism from that of Frey (1962), who reported that the sensations he was studying seemed to be associated with noise riding (high-passed) on each of his radar pulses.

To verify their idea quantitatively, FF74 used a hydrophone to measure an acoustic response to 27 $\mu s$ pulses of 2.4 GHz EMR in a somewhat larger than head-sized container of pure water above, at, and below the 4 C temperature at which liquid water reaches its maximum density. They found that the acoustic wave amplitude declined to zero as the temperature was lowered to 4 C, and a nonzero amplitude returned with phase reversed as the temperature was lowered below 4 C.

The null result at 4 C can not be explained any obvious way except by a superposition of two scalar acoustic waves of equal amplitude but opposite phase.  The null point for their sound wave being at the same 4 C as the turning point of water density, FF74 concluded that their radar pulses were delivering temperature pulses to the water.  The reversal in phase strongly supported this conclusion.

To argue further in favor of the concept of specifically <u>thermo</u>acoustic such waves, FF74 cited calculations and data by Gournay (1966) which indicated that a high-power optical pulse could cause acoustic waves in water by thermal means.

FF74 postulated a temperature increase per pulse of at most $10^{-5}$ K.  This estimate is consistent with acoustic calculations and results by Bacon, *et al* (1999).  Let us assume a human head 20 cm in longest dimension, and an auditory system sensitive to linear displacements of some $10^{-11}$ m amplitude (Bekesy, 1948; Bekesy and Rosenblith, 1951).  Then, for a fundamental-mode, lossless vibration at absolute threshold, this amount of temperature increase would imply a linear coefficient of thermal expansion, $X$, given by, $10^{-11} < (20/100) \cdot 10^{-5} X$.  Higher-order (odd) modes would cause far smaller acoustic contributions, no more than a dB or so, and will be ignored.  So, approximately, a threshold sensation would require $X > 5 \cdot 10^{-6} / K$.  The linear expansion coefficient $X_{H_2O}$ for water is about $3.7 \cdot 10^{-4} / K$ at body temperature (Chaplin, 2002), and common materials differ in linear coefficient on the order of $10^{-5} / K$, so the idea definitely seems consistent quantitatively with known properties of the human auditory system.  The sequence of events is sketched (not to scale) in Fig. 6.

Accepting Foster and Finch's estimated maximum temperature increase per pulse, the thermoacoustic effect from each temporally distinguishable pulse should contribute a click no more than about $(X_{H_2O} / X)^2 \cong (10^{-4} / 10^{-6})^2 \cong 40$ dB above auditory absolute detection threshold; this is about the level of the softest whispered speech.  EMR or acoustic wave diffraction, if present, could alter this estimate by perhaps an order of magnitude.

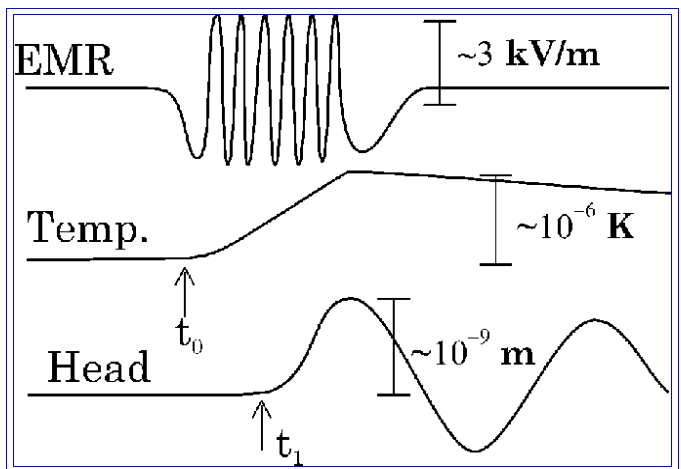

**Figure 6.  Sketch of events in the Foster and Finch (FF74) hypothesis.  No scale.  The EMR pulse arrives at the head at time $t_0$; the head begins to oscillate at time $t_1$.**

Support for a cochlear source of auditory response both to EMR and to ordinary sound stimulation was reported by Lebovitz and Seaman (1977), who recorded single-cell responses in the cat auditory nerve.  They concluded at the time that the FF74 hypothesis was consistent with their results.  For cells responding comparably to EMR and sound stimuli, many of the Lebovitz and Seaman EMR pulses were calculated to be equivalent to a sound click 60 to 70 dB above (human) absolute threshold.

   The problem here is that 60 to 70 dB seems orders of magnitude too high for the thermoacoustic hypothesis above, given the small size of the cat skull.   The 60 dB figure would require a cat skull bigger than an elephant's.   Furthermore, the problematic Lebovitz and Seaman data don't seem explainable in terms of special properties of the cat auditory system, because data by Cain and Rissmann (1978) indicate that cochlear sensitivity to sound or EMR is not much different for cats than for humans.   Seaman and Lebovitz (1987) subsequently decided that the FF74 thermoacoustic hypothesis was not tenable for cats, but their changed conclusion essentially was ignored by thermoacoustic proponents.

   There are other ways in which the thermoacoustic idea does not seem to be a complete explanation of microwave hearing.   First, purely theoretically, Gournay (1966) assumed that the thermal contribution would be the only one in his experiment, and he calculated optical-acoustic wave efficiency from that.   There is no easy way to test such an assumption from his study or his calculations.   FF74 seem merely to have repeated back Gournay's assumption to support their own conclusion.

   Also, even if the Gournay assumption was entirely correct, the FF74 extrapolation from Gournay's visible light to frequencies more than $10^5$ lower should be substantiated more directly by data in the microwave range.   The Gournay experiment was performed at frequencies far above the empirical first resonance point of water, which is about 30 GHz -- so far above, that hydrogen bonds in water could not possibly be stretching or bending in synchrony with the EMR.   The FF74 experiment was performed at frequencies so far below this point that the main electromagnetic response of the water would be in dipole-mediated, bending changes in hydrogen bonding (Chaplin, 2002).   Thus, the mechanism of response of water to Gournay's *versus* Foster and Finch's EMR had to be quite different, whether "thermal" or not.   The same argument would apply to subsequent findings by Olsen and Lin (1983), who demonstrated the presence of acoustic waves caused by radar pulses applied to animal tissue.

## A Structural Acoustic Alternative to Foster and Finch

   The FF74 demonstration can be explained by a nonthermal acoustic mechanism.   Recalling the water structure discussion above, suppose each EMR pulse in the FF74 frequency and power range randomly merely caused some small fraction of all dipoles on the boundaries of the ES icosahedra to rotate their hydrogen bonds enough to buckle inward, which is the equivalent of shifting some ES to CS.   Little or no additional random motion would be involved (although the new CS icosahedra would store some new energy), so there would not be any significant temperature pulse.   Buckled molecules on the surfaces of the icosahedra would disconnect their hydrogen bonds from other, bounding structures; these other structures would be to some extent fragments, which we know increase in number with temperature.

There would be three different, temperature-parameterized cases:

- Below 4 C, an EMR pulse would be converted to a ES → CS density-increase pulse unrelated to any immediate change in the heat content of the water.

- At 4 C, the ES → CS density increase would be cancelled by the density decrease caused by newly freed fragments.

- Above 4 C, CS+fragments would dominate over ES, so a pulse would deliver a density-decrease pulse unrelated to any immediate change in heat.

The point here is just that the peculiar density maximum of liquid water at 4 C results from its special intermolecular structure, not from temperature. Of course, whatever the mechanism, eventually most of the EMR energy delivered to the water and not converted to an acoustic wave will remain in the water as heat, raising slightly the water temperature.

We prefer to reinterpret the FF74 observations using the icosahedral ES – CS equilibrium espoused by Chaplin (2002); however, any theory accounting for the 4 C density maximum of water would seem indifferently to support either a temperature pulse or a structural pulse as the cause of the acoustic wave demonstrated by FF74.

Thus, the dodecahedral (over 0.5 nm diameter) or smaller water structures measured by Miyazaki, *et al* (2004) or Shin, *et al* (2004) can not strengthen the FF74 hypothesis, whereas the many FF74 inconsistencies with the observed data, expanded upon below, tend to refute it except possibly at very low, "nonthermal" levels.

### Structural Acoustic Hypothesis Tests

The preceding structural explanation of an EMR-created acoustical wave is meant to show that the FF74 pure-water conclusion is not necessarily correct, and that another explanation is tenable. However, the explanation also implies testable hypotheses about microwaves and water:

One test would be to examine GHz-range EMR pulses on a fine time scale.

According to FF74, the water is delivered thermal energy, some of which may be carried off by an acoustic wave. The temperature rise during a pulse is simultaneous with the density change launching the acoustic wave.

Under a structural hypothesis, as the EMR pulse continues, structural change accumulates, increasing (or decreasing) water density and thus launching the acoustic wave.

Looking at finely resolved temporal details, and assuming a water temperature away from 4 C, a structural explanation should be favored by (*a*) a density change in the water preceding the first temperature increase; or, (*b*) a dip in water temperature during the EMR pulse.

Another test would be to measure the total heat delivered to the water at 4 C and at nearby temperatures.

The FF74 hypothesis assumes that the water should show the same pulsed temperature increase at 4 C as at nearby temperatures.  At temperatures other than 4 C, the acoustic wave must carry away some energy.  So, the total heat delivered by the pulse should be <u>equal or less</u> at 4 C than at nearby temperatures, because the given temperature pulse should be attainable at 4 C with a minimum of heat input.

In a structural hypothesis, the density change follows structural energy absorption; almost all EMR energy which will be absorbed is absorbed into structural change before it becomes a density change.  Thus, at 4 C, the energy which otherwise would go into the acoustic wave remains in the water: An EMR pulse should deliver <u>more</u> total heat to the water at 4 C than at slightly different temperatures.

### A Nonacoustic, Nonthermal Direct Hearing Hypothesis

The present author's preferred explanation of microwave hearing is that the EMR causing microwave hearing interacts directly with the cochlea, for example electrically or mechanically.  In doing so, this EMR directly would affect sensory elements of the cochlea, possibly the individual hairs of the hair cells.  In turn, these cells, or local regions nearby, would propagate the effect through the rest of the auditory system, ultimately causing auditory nerve impulses, as would have an acoustic stimulus.

However, EMR has no anatomical basis for coherently or sequentially moving larger structures of the cochlea, such as the basilar or tectorial membrane, in the way an acoustic stimulus would.  There would be no travelling wave or place specificity as would occur for a real sound; so, EMR could not produce a cochlear response equivalent to that caused by an acoustic stimulus either by bone conduction or by air conduction.  Therefore, direct cochlear stimulation by EMR could not correspond to a physically possible acoustic stimulus.  The EMR in effect would be uncorrelated -- randomized -- with respect to any physically possible overall cochlear response to a sound originating in the real world.

This would be why, in microwave hearing, the listener can not perceive the loudness or pitch of the "sound" very well.  Also, localization of the source of the apparent sound would be poor, as is reported for microwave hearing, because the stimuli in the two cochleae, processed by the auditory system as though acoustic, in general would be random and incoherent with respect to any possible acoustic stimulus.  This last might be tested by irradiating the two temporal regions of the head with EMR pulses differing by a time delay equivalent to a localization delay known effective for normal human audition.

### Microwaves as Auditory Stimuli

Obviously, in microwave hearing as in normal hearing, morse code or other click-encodings are possible; and, chirpy tones have been reported (Guy, *et al*, 1975); but, transmission of speech or music has been found difficult (Guy, e*t al*, 1975, *Discussion*).  Early reports of intelligible speech by microwave (see Justesen, 1974) have not been confirmed and probably were mistaken.

Under an acoustic hypothesis, microwave hearing should permit speech or music transmission by microwave, simply by modulating pulse amplitudes (envelopes) and timing so as to synthesize appropriate bone-conducted acoustic patterns in the human head.  A clever EMR transmitter system under computer control should be able to synthesize sound waves in the head from acoustic pulses, because the human head is on the large side of the acoustic diffractive range, as mentioned above.

Under a direct EMR hypothesis, microwave hearing would not be expected to permit modulation to speech or music, because temporally modulated random (direct EMR) responses would not be expected to be anything but further randomized in space and time in the cochlea.  One might hope that a complicated spatial, diffraction-pattern, modulation of microwave pulses might be generated, perhaps under computer control, so as to interfere in the cochlea to synthesize time-changing patterns the same as acoustic travelling waves corresponding to speech or music patterns.  The difficulty here is that the cochlea is only a few mm in size, and microwave hearing in humans has not been reported for EMR frequency content above about 10 GHz (Foster, 2000), implying a wavelength of several cm.  The cochlea is on the small side of the diffractive range for microwaves even at 10 GHz.  So, sound-synthesizing spatial patterns could not be resolved using EMR at wavelengths known capable of producing microwave hearing sensations.

A possibility under a direct EMR hypothesis would be to create auditory effects depending on widely spaced clicks, beyond the cochlear summation interval for EMR.  For acoustic stimuli, this interval is around 200 $\mu$s  (Bekesy and Rosenblith, 1951).  This would suggest the possibility of transmitting very low-pitched speech or music, by modulating EMR pulses delivered at, say, one or two kHz.  The voice quality would be very low, like speaking through a kazoo.

Finally, if an acoustic effect indeed were present in microwave hearing, in addition to a direct EMR effect, possibly some way of producing the acoustic effect in the absence of any direct EMR effect might make microwave transmission of speech or music feasible in very quiet surroundings.

### The Underlying Thermal Problem

There would seem to be a need for improved understanding of the acoustic effects on cold water of pulsed EMR in the GHz microwave range.  Whatever these effects might be, though, a temperature-induced acoustic wave only 40 dB above absolute threshold would not seem loud enough to explain microwave hearing.

Moreover, Sommer and von Gierke (1964) were able to produce microwave hearing sensations with audio-modulated electric fields, thus ruling out any possible thermal input.  None of the spatial characteristics of diffractive-range EMR were present, as also pointed out by Guy, *et al* (1975), which perhaps might account for the much higher hearing threshold than was obtained by Frey (1962), who used radar pulses.

Any acoustic hypothesis depends on bone-conducted vibrations of the head and would predict considerably higher thresholds in small-headed animals than in humans. On the other hand, a direct-action hypothesis would predict no difference, because the action would be on some commonality of the cochlear sensorium, which is about the same for all mammals. Small-animal microwave hearing thresholds have been measured equal to human thresholds within about a factor of two (Cain & Rissmann, 1978; see also Seaman & Lebovitz, 1987), so the evidence seems to favor a direct EMR action.

There must be a mechanism for microwave hearing not related to the thermoacoustic one described in FF74. The thermoacoustic explanation predominates in the review by Chou, *et al* (1982), in Erwin (1988), and in other similar papers. Swerdlow (2003) explicitly adopts a "thermal" view of microwave hearing. Many authors, even through the years following 2010, seem to have picked up the idea of a thermal mechanism from FF74, to have fixated upon this one explanation, and to have ignored other, possibly more consistent, explanations. Even worse, the thermal approach to microwave hearing has been applied indiscriminately to discredit the existence of any microwave biological effect not caused by conversion of EMR to heat. Conceptual difficulties thus have impeded progress in this field and often have led to contradictory and irreproducible findings.

We now pass from microwave hearing proper, to unrelated concepts often invoked pro or con the FF74 hypothesis.

## *Spherical Assumptions*

In describing the microscopic level of response of living organisms to EMR, a "cell" of two or more concentric spheres, a convenience in solving the the Laplace equation, often has been assumed. This is quite unrealistic but would be justified because of the limited computational resources which were available in past years.

Assuming a sphere for a cell is an oversimplification reminiscent of the one that microwaves induce only heat, because heat can be calculated and measured easily.

The spherical cell approach was used by Drago, *et al* (1984): They simulated a multiple-shell spherical model and ostensibly found a sensitivity of such constructs to 100 Hz EMR. Foster (2000) discusses the spherical model at length, raising some partially correct questions about the Drago, *et al* conclusion. To avoid confusion of the reader, we should emphasize that the Drago, *et al* study is being cited here because of the cellular model, not because of evidence of a 100 Hz biological interaction.

Foster (2000) concluded from a spherical-cell model that the response to microwaves would be negligible. However, this conclusion seems incorrect for humans, if not for the Drago, *et al* simulation model. The underlying problem is that spherical symmetry introduces field cancellations generally inapplicable to living, multicellular organisms. The sphere cancels mathematical complexities of the problem, but it cancels biological effects, too. The only spherical cells in the human body are fat cells or perhaps some kinds of blood cell; modelled as ellipsoids, sensory or motor nerve cells would have eccentricities on the order of $10^4$ or more. We introduce a new approach below, showing how hyperpolarization even of a sphere could occur; but, for now, we address the spherical symmetry as a hindrance to be overcome.

Because of the unrealistic symmetry, we are led to assume here that, in general, spherical models must be irrelevant to the question of nonthermal interaction with human cells. However, perhaps a spherical model might be adapted to the shapes of actual living cells:

Consider a fairly straight cell process with diameter of $10^{-5}$ m and length of $10^{-1}$ m; many human nerve cells would fit this. Given the eccentricity of $10^4$, one might approximate such a process by a chain of $10^4$ of the objectionable spheres, each of twice the cell radius to ameliorate surface vector cancellation because of symmetry. Using Eq. (12) of Foster (2000), the capacitivity to conductivity ratio would increase lengthwise by a factor of about $10^4$; radially, it would decrease by $2/2^2 \cong 1/2$. The result would be a time-constant of maybe 0.1 $\mu s$ perpendicular to the long direction, but one of about 1 ms in a lengthwise direction. The membrane potentials in Fig. 6 of Foster (2000), then, would vary by some factor between 1 and $10^4$, depending on direction of the EMR. So, one would expect a membrane potential between $10^{-6}$ and $10^{-2}$ V for a 1 V/m incident $E$ field. Pulsed ultrawide band (UWB) fields may be expected to reach several V/m, so EMR-induced membrane potentials several times the normal neuronal resting levels of $-70$ mV should be possible. This conclusion, using the same basic model as does Foster (2000), but without the symmetry, is opposite to the one reached by Foster.

A more recent simulation of a small number of spherical cells in close proximity of one another (Sebastian, *et al*, 2001) tends to confirm these rough calculations for chained spherical elements.

## *Time-Domain Patterns*

As mentioned above, reflections of incident EMR wave components will be found in the diffractive region everywhere near organ boundaries, especially where muscle and either bone or skin are close. For GHz-range microwaves, the medium always being absorptive, the reflected wave amplitude will be smaller than the incident, creating an oscillating, standing field with a net flow of energy into the tissue component, the polarity depending on local phases (Durney, *et al*, 1999, 3.2.12). The geometry of the resultant field essentially will be unrelated to incident frequency, except for being multiple half-wavelengths in size, but it will be fairly closely related to the geometry and tissue properties of the organism. If the incident field is moving in phase, or if the organism is moving in space, the standing patterns may be transitory, persisting perhaps for milliseconds or seconds. Under proper conditions, the time-change of the field amplitudes should be able to impose centimeter-scale, changing, diffractive spatial patterns on nerve cell receptive processes, which should cause sensory responses.

It is not our purpose to present a fully developed theory here; we wish only to lay conceptual groundwork for correct understanding. Still, let us examine the effect of these patterns in some detail, showing just how they would represent changing patterns of membrane polarization:

Consider any membrane fragment large enough in area to contain several of the dipoles represented in the previous figures.   As shown in Fig. 7 below, the membrane polarization may be represented as an electric potential difference, *V*, and viewed as a dipole.   This dipole creates an *E* field perpendicular to the membrane surfaces which tends to cancel the field created by the ionic concentration differences on opposite sides of the membrane.   Although other dipoles will be present in the polar fractions of membrane molecules, this dipole is the one in the membrane bulk or pores which cancels the applied voltage; it is what causes the membrane permittivity to exceed greatly the vacuum permittivity.   The same permittivity increase is responsible for the increase in capacitance of an air-gap capacitor, when an insulating sheet is positioned between the plates.

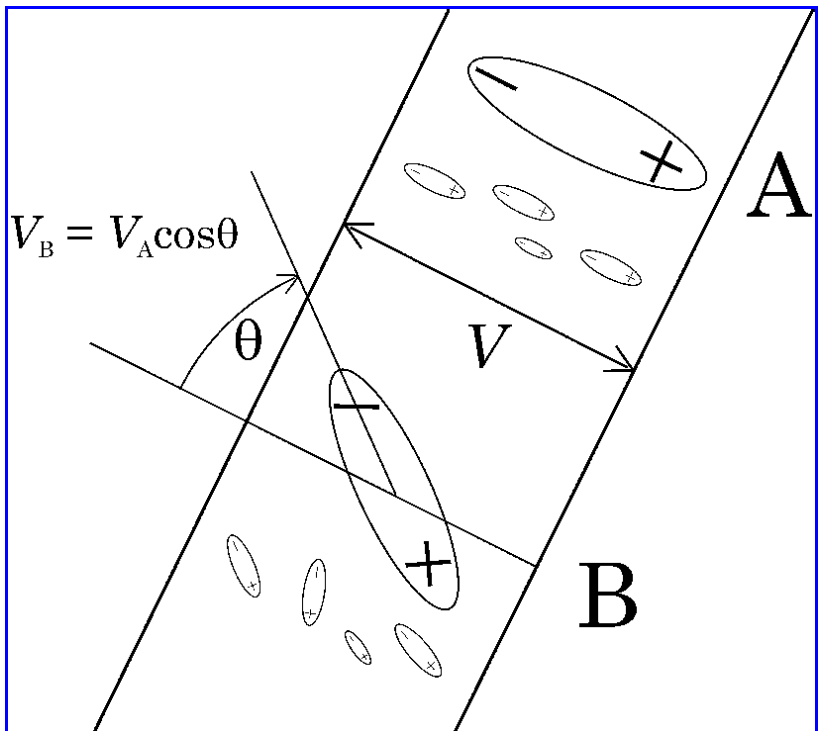

**Figure 7.  Hyperpolarization of a membrane by EMR. The large dipole represents a sum of smaller ones in the membrane which tend to cancel the membrane ionic potential.   The dipole potential V across the membrane in A is less in B after rotation of the membrane dipoles over some angle, θ, causing a net increase in total membrane potential.**

In the upper part of Fig. 7, labelled "A", the polarization in the absence of EMR is represented as because of a single dipole.   The actual structure of the membrane would suggest that this dipole should be the resultant of the sum of divers smaller ones also shown schematically in the figure.   Thus, the measured voltage $V_T$ across the membrane may be represented as a sum of the ionic potential $V_I$ and an opposing polar potential, $V$ ; therefore,  $V_T = V_I - V$.   If $V$ is decreased, the membrane potential $V_T$ will increase, so the membrane will hyperpolarize.

If we assume that the EMR frequency is reasonably low enough, some or all of the dipole(s) in the membrane will rotate almost in phase with the EMR, because a reasonably low frequency will allow the dipoles enough time to relax in the opposing forces of (*a*) the incident EMR and (*b*) the structural, or perhaps functionally active, bonds internal to the membrane.   Such a rotation angle is shown in part "B" of Fig. 7.

If we also assume that biological events on the scale of such a membrane fragment occur slowly when compared with the EMR frequency, the biology will see not the instantaneous dipole fields, but rather their time average.

Both of these assumptions will be fulfilled for EMR in the GHz range below about 10 GHz.

The rest is trivial: We represent the time averaging of the dipole potential by the absolute value of the time-average angle, which we call $|\theta|$; the value of $|\theta|$ necessarily must exceed 0, if any rotation occurs at all. The cosine of an angle always is less than 1, unless that angle is 0. Of course, rotation in either direction, clockwise as shown in Fig. 7, or counterclockwise, has the same effect of decreasing $V$, thus increasing the total potential across the membrane; this is the reason for the absolute value. Therefore, if incident EMR causes rotation of membrane dipoles by $|\theta| > 0$, it must hyperpolarize the membrane. This rotation eventually may generate heat in the membrane or surrounding tissue or fluid; however, such heat clearly will be incidental and will not be related directly to the hyperpolarization.

Note also that even for a membrane surrounding a spherical or (more realistic) circular-cylindrical inner space, the symmetry will <u>not</u> cancel this hyperpolarizing effect; rather the hyperpolarization will sum over the entire area of the membrane, provided only that the wavelength of the incident EMR considerably exceed the diameter of the inner space.

Finally, note that there will be inertial forces, and mechanical (elastic) restoring forces, not only electrical ones, that determine the extent of the dipole rotation in an incident EMR beam. The numbers can be deceptive: True, the transmembrane $E$ field because of a resting potential of 100 mV across a membrane 10 nm wide would be about $10^7$ V/m; this might suggest substantial, essentially immovable, rigidity of the dipoles. However, as pointed out in the discussion above ("EMR and the Human Body"), charges in body fluids during irradiation by EMR will be rotated as dipoles, or displaced, to cancel the forces exerted by EMR-induced potentials. So, almost any EMR $E$ field superposing a potential in the millivolt range would be expected to be capable of strengthening an otherwise resting membrane field: The impedance of charge movement would be across the same membrane, of the same width, as was sustaining the comparable resting potential. The enormous, microscopic $E$ field is only apparent and easily is modified by comparably microscopic events.

It should be kept in mind that hydrophobic dipoles exist in the bulk of the membrane, while membrane hydrophilic dipoles exist only at the membrane surface(s) and in the channels which control membrane functionality. So long as the average dipole moment differs in the membrane (or channel) from the fluid media surrounding it, microwaves will cause a voltage difference across the membrane, although the actual difference may vary from that to be estimated below.

Now, arriving at an opportunity for quantification, first let us look into the question of the amplitude of dipole rotation in terms of mechanical characteristics of the dipole. We want, above all, to be sure that some rotation can occur at a reasonably low EMR power density. Even a few percent change in membrane potential probably would be enough of a hyperpolarization in a sensory nerve dendrite or end-organ to initiate a response of some kind -- a response well correlated with the presence of the EMR.  It isn't hard to show this, at least within an order of magnitude:

Assume a membrane width $d$ and a dipole molecule with charge separation $2a$ and thus with electric dipole moment $2qa$ ($q$ = 1 unit of electric charge).  Let the dipole molecule mass be $m$.  Also assume a dipole molecule moment of inertia of $I = m(2a)^2/\kappa$, with $2.5 \leq \kappa \leq 12$ a shape parameter varying from spherical up to that of a thin rod.  We shall at first ignore the applied EMR field strength and assume that the angular frequency $\omega$ of dipole rotation will be equal to that of the external EMR.  This is meant to allow some estimate of the dipole rotational amplitude solely in terms of its mechanical characteristics.

From the above, then, the energy state of such a dipole in the membrane field $E_m$ may be written as

$$\frac{1}{2} I\omega^2 = 2qaE_m(1 - \cos\theta). \tag{3}$$

So, to estimate the average dipole displacement angle $\theta$, we may use the formula,

$$\cos\theta = 1 - \frac{ma\omega^2}{\kappa q E_m}, \tag{4}$$

Let's assume that the dipole shape is somewhat ellipsoidal, say $\kappa = 4$, and thus approximate the volume of the dipole molecule by $V = \pi a^3$.  We know that the density of water is 1 $g/cm^3$, which is the same as $10^3$ $kg/m^3$.  Assuming the effective density of a typical membrane dipole molecule to be $3/4$ that of water, we shall use 750 $kg/m^3$ as the density of such a dipole.  Eq. (4) then becomes,

$$\cos\theta = 1 - \frac{(750 \cdot \pi a^3)a\omega^2}{4(1.6 \cdot 10^{-19})E_m} \cong 1 - \frac{3.7 \cdot 10^{21} \cdot a^4 \omega^2}{E_m}, \tag{5}$$

For membrane width $d$ = 10 nm and membrane potential of 70 mV, we have the field $E_m = 7 \cdot 10^6$ V/m.  With EMR of 1 GHz, a reasonable range of values for the relation of the angle $\theta$ to the half-length of the dipole given by Eq. (5) is plotted in Fig. 8 below.

Looking at Fig. 8 below, there is a significant amplitude of rotation for a dipole of half-length anywhere over about 1 nm, so the idea at least of some minimal membrane hyperpolarization by dipole rotation is supported on this account.

Naturally, rotation over large angles, say $|\theta| > \pi/4$ would imply either broken bonds, and thus a damaged membrane, or rotation in a fluid; so, we shall proceed further on the assumption of a rather small angle $\theta$.

The preceding calculation obviously leaves much to be desired: Clearly, for the cosine to be defined by Eq. (4), we must have values $0 \leq ma\omega^2 \leq 2 \cdot \kappa q E_m$, so there is missing at least a proper description of the effect of changing the frequency.

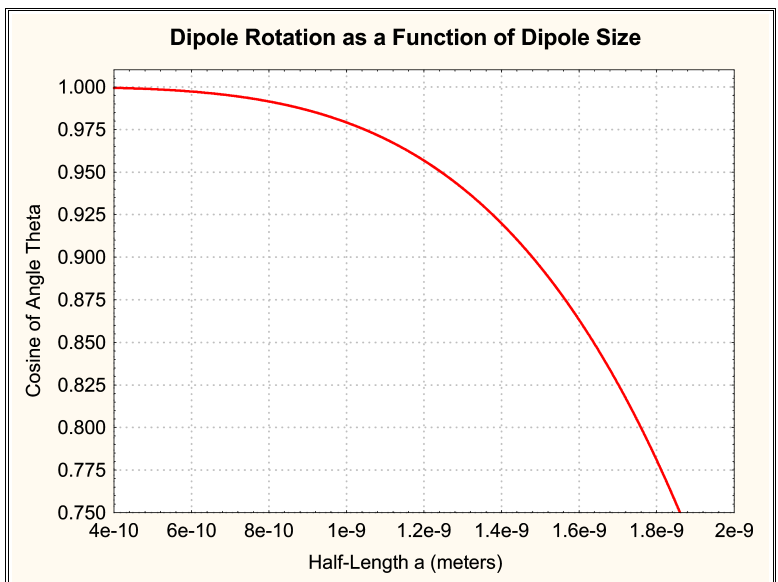

**Figure 8. Mechanical response to 1 GHz EMR of a somewhat ellipsoidal dipole, according to text Eq. (5). The resting membrane electric field is assumed to be $7 \times 10^6$ V/m.**

For example, to criticize the preceding calculations, an increased frequency perhaps should force segmented rotation, reducing the length and mass of the average dipole able to follow the phase of the applied EMR. Not unrelated to this shortcoming, the elastic, Hooke-like restoring forces of the dipole-matrix bonds are ignored in Eq. (5); for a given amplitude of EMR, these would reduce the average angle $\theta$. Such restoring forces by definition are assumed small compared with those keeping the dipole together. A related difficulty in further quantification is in the local change of $E_m$ to displacement changes in the membrane charges caused by coherent rotation of numerous dipoles. The value of $E_m$ in Eq. (3) would be increased in a region of membrane populated by numerous dipoles, all rotated to some phase angle greater than 0, thus decreasing the rotation angle below that expected from Eq. (5) or from the equations below.

With all this in mind, let us look at the problem in terms of the EMR field strength. In the previous calculation, we took for granted that the EMR could rotate the dipole *because of* its mass, or, more precisely, because of its rotational kinetic energy. However, $F = ma$ implies $a = F/m$; and, clearly, from the left side of Eq. (3), the driving EMR must accelerate the dipole constantly. A very massive dipole hardly would rotate at all in the EMR field, meaning that $|\theta| \sim |\phi|/m$ will go to zero as dipole mass $m \to \infty$.

So, ignoring nonlinearities and other complications, let us examine this problem more closely by assuming some minimal dipole rotational amplitude and rewriting Eq. (3), replacing the mechanical energy with an expresssion for the required driving EMR energy, $U_{EMR}$, in number of photons:

$$U_{EMR} \equiv n\hbar\omega = 2 \cdot qaE_m(1 - \cos\theta) \quad \Rightarrow \quad n = \frac{2}{\hbar\omega} \cdot qaE_m(1 - \cos\theta). \tag{6}$$

Then, again for the same $\omega$ and $E_m$ as above, the required number of microwave photons $n$ is given by,

$$n = 3.4 \cdot 10^{12} a(1 - \cos\theta). \tag{7}$$

This result is plotted as a set of contours in Fig. 9.

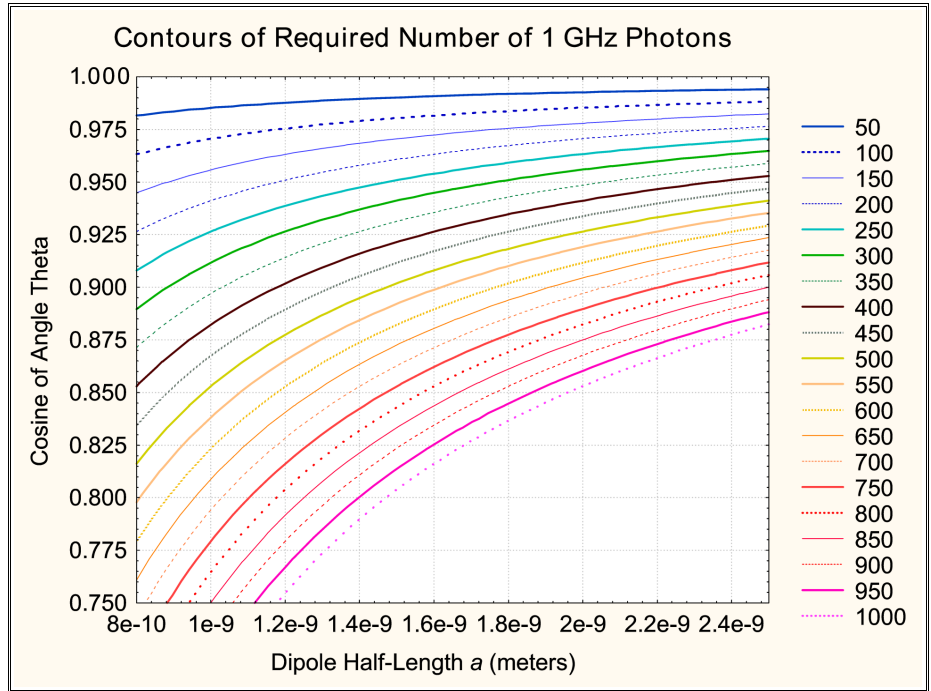

**Figure 9.  Number of 1 GHz photons required to rotate a dipole of half-length $a$ to an angle of cosine $\theta$ in the assumed membrane electric field, as given by text Eq. (7).**

Looking at the vertical axis of Fig. 9, anywhere there are contours would give a rotation angle probably capable of causing a neural or muscular hyperpolarization causing some sort of noticeable effect.   However, we prefer small angles with cosine no less than, say, 0.9; so, we are restricted to about the upper third of the figure.

Looking at the horizontal axis and comparing Fig. (8), any dipole length at all in Fig. (9) probably would be consistent with our assumptions.   So, something around 100 photons per dipole would be required for a 2% (1.2 degree) rotation.

We are interested in the energy delivered by the EMR per unit area, so we should restate Eq. (6) or Eq. (7) in terms of membrane area.   Taking into account some ellipticity in the dipole shape, the result is,

$$U_{EMR} = \frac{3.4 \cdot 10^{12} \hbar\omega \cdot a(1-\cos\theta)}{(3/4)\pi a^2} \equiv \frac{2 \cdot qaE_m(1-\cos\theta)}{(3/4)\pi a^2} \cong 9.5 \cdot 10^{-13}(1-\cos\theta)/a\ ; \qquad (8)$$

and, this is plotted in Fig. 10.

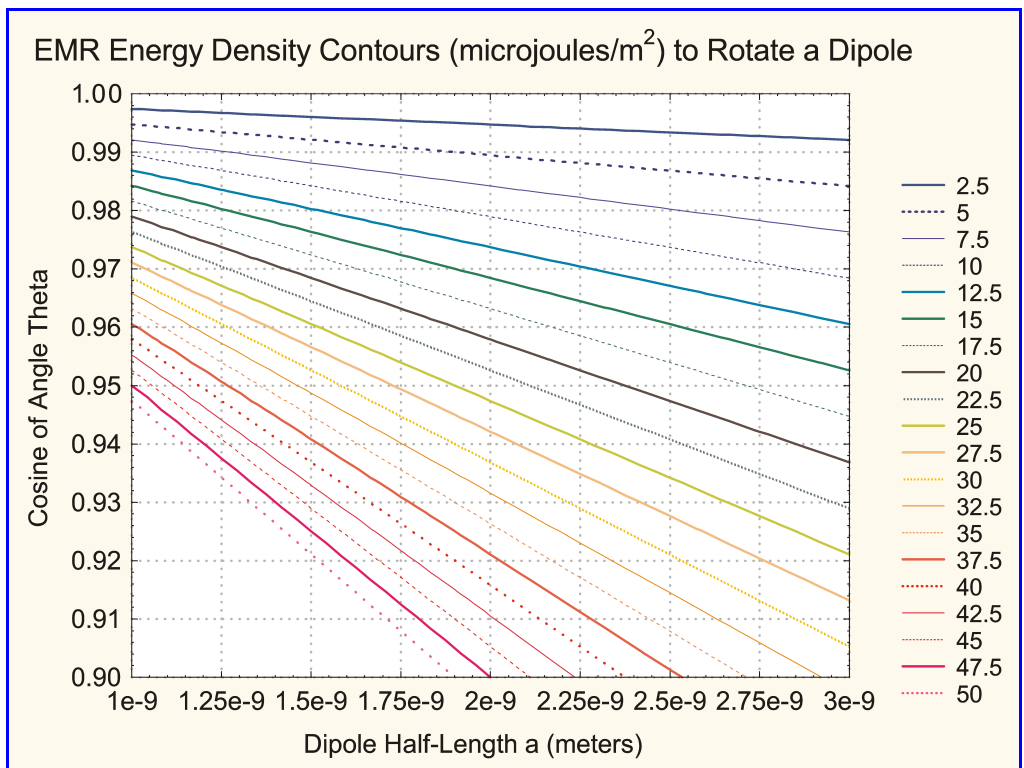

**Figure 10. Energy per unit area required to rotate an ellipsoidal dipole in a membrane field of $7 \times 10^6$ V/m by 1 GHz EMR, according to text Eq. (8).**

To examine this result more closely, let's pick a dipole 4 nm long (2 nm half-length) in Fig. (10) above. The energy density required, from the figure, would be under 15 $\mu J/m^2$.

During successive EMR cycles, a large fraction of the energy required for rotation would be returned to the system by elastic response of the bonds holding the dipole in its membrane matrix, but even requiring that the EMR supply as much as 15 $\mu J/m^2$ per cycle, at 1 GHz the irradiant power would be about 15,000 $W/m^2$ or 1.5 $W/cm^2$. This could be within safe thermal exposure limits, by current national or international rules.

One final thing: Our calculations here are based on a resting membrane potential and its accompanying very high local field of some $7 \cdot 10^6$ V/m. However, during normal function, during ordinary slow depolarization or passage of an action potential, a nerve or muscle membrane may develop a much lower field, even one close to zero V/m. In these occasional conditions, incident EMR would be expected to rotate membrane dipoles over far greater angles than those calculated above. Thus, active muscle or nerve cells would be expected to be affected more by EMR than those in a resting state.   So, our calculations above are for a very worst-case membrane state -- very conservative in terms of demonstrating the plausibility of membrane polarization as an effect of EMR.

Although this entire formulation leaves plenty of room for refinement, we conclude that the results are in the range of the possible; and, the idea of polarization change by EMR dipole rotation seems tenable quantitatively, the main uncertainty depending on computational assumptions mentioned above.

This polarization change implies something very interesting in terms of the time-domain patterns of irradiation:  The human skin is innervated with "C" sensory nerve fibers which mediate various tactile sensations, notably pain.   These sensory processes are unmyelinated, very small in diameter, and very close to the surface of the skin.   The small diameter and superficial location implies that they would require a relatively low power input to drive their membrane potentials to follow the location on the skin of moving EMR amplitude nodes or antinodes.   One would expect this driving to establish spatial patterns of hyperpolarization on these sensory fibers.

It seems reasonable that dynamically changing spatial patterns of centimeter-scale standing microwave components on these "C" fibers would provide a theoretical basis for use of UWB or other pulsed radar transmitters to harass living humans and animals.   However, sensations possibly caused by "C" fiber irradiation have not been studied systematically, to the present author's knowledge.

## *Thermal Noise and EMR*

A common argument is that the random velocity of ions in solution would be some orders of magnitude greater than the velocity added to one of them even by a field of, say, 1 kV/m; therefore, all EMR effect on charge carriers in a cell or cell membrane would be drowned out in thermal noise.

Boltzmann's formula may be used to relate heat energy to the kinetic energy of molecules of mass $m$, by setting $kT \propto mv^2$.  The energy of a microwave photon at 1 GHz is just $E = \hbar\omega \cong 4$ $\mu$eV ; from $\frac{3}{2}kT$, the energy of an elementary, ideal gas, particle at 300 K is about 40 meV; so, apparently, such a photon could contribute no more than about $1/10,000$ the energy of the (somehow conceived as ambient) temperature.   This energy actually is an expectancy, and the contribution of temperature to velocity in any specific direction will be just $\frac{1}{2}kT$.  Any way, the velocity added to an absorbing element by such a photon would be expected to be something around $1/100$ that of any one random, thermal collision.   Thermal noise also is invoked by Foster (2000).   However, there are serious pitfalls to avoid:

First, the extracellular fluids are not just filled with small ions, but also with macromolecular assemblages (Regan, 1972, p. 1) with much greater organization than the "ideal gas" on which Boltzmann's formula is based.   In addition, water itself is far from a liquid of independently moving water molecules, as has been explained above. At a given temperature, then, the thermal velocities of biologically important assemblages will be far less than the ones presented, for example, in Foster (2000). Much of the fluid thermal energy will be internal to the assemblages and will not be available to add noise to EMR-membrane interactions. Schulten (2000, p. 61), for example, remarks that "[proteins and RNA] work at physiological temperatures and thus experience thermal motion, yet their function, which requires correct alignment of parts and steering of reactions, is executed with precision."

Second, the quantification based on Boltzmann's ideal gas becomes even more inaccurate near a boundary surface or membrane, away from the central bulk of a fluid. Near any boundary in a living cell, there will be a partially screening region of high molecular correlation which separates the membrane proteins from the somewhat more gas-like bulk fluid. Drago, *et al* (1984) represented this screening by the "bound water" layer in their spherical model; it is a general property even of pure liquid water and is discussed also in Chaplin (2002). Molecular size and screening correlation both reduce the velocity of random thermal motion.

The effect of a photon on a large, relatively slowly moving structure does not fit the simple Boltzmann analysis above. For a 25 kiloDalton membrane segment or large protein molecule, the thermal speed would be just $1/\sqrt{25,000} \cong 1/150$ that of a hydrogen atom at the same temperature, making the expected velocity added to a constituent hydrogen atom by an absorbed 1 GHz microwave photon greater than that added by the average thermal collision.

Finally, the effect of an *E* field on a fluid is to cause a drift in the mean position of the charge in the bulk by rotating polar elements or, at low frequencies, by displacement of the mean charge. The field does not somehow have to segregate and accelerate a subpopulation of the charge carriers against a thermally randomizing background. This last highlights an underlying conceptual problem with the thermal argument: Live membranes, as discussed above, have no need for the functionality of the P-N junction of a diode to convert incident EMR into work. Such membranes operate by polarization, not by rectification. The model of a radio receiver in a living organism, making it work by detecting signals against a background of membrane noise, is completely wrong.

A reader disagreeing with the *E*-field drift statement above might try a simple experiment: Take a 9 V transistor-radio battery (approximately 10 V/cm spacing of the terminals; or, 1 kV/m) and touch both terminals briefly to the tongue; see whether random molecular motion from body heat can drown out the current flow! Of course, such a battery can supply chemically-powered current to fulfill Ohm's law of the tongue; an *E*-field in space would be loaded down substantially -- an issue completely overlooked by the random-noise argument. Nevertheless, a transmitter producing a continuous microwave output with amplitude 1 kV/m might be unhealthy for anyone nearby; if it could deliver the power, for example in the reactive near field, to maintain 1 kV/m in the bulk of a nearby human body, it might well be deadly. A cell phone held to the ear puts the head in the reactive near field of the phone's antenna; however, the field would not be anywhere near 10 V/cm (1 kV/m).

Random thermal motion simply is not relevant to microwave *E* field effects on ions in solution, unless the discussion has been limited to one concerning microwave ovens. There are at least two good reasons that a UWB microwave transmitter does not produce painful electrolysis in everyone nearby, and they have nothing to do with thermal randomization: (*a*) <u>Duty cycle</u>: Typical duty cycle for a pulsed UWB transmitter with peak field of 10 V/cm would be, say, 0.001. So, the (rectified) average electrolyzing potential would be about equivalent to that of a few mV/cm continuous field, producing little or no noticeable thermal or electrolytic effect. (*b*) <u>Distance</u>: Intensity typically falls off as the inverse square of distance. A small antenna transmitting an average of 1 W typically will be nondirectional. Even in a device held against the head, no more than about 40% of the power radiated (0.4 W) will be intercepted by the user's head (Li, *et al*, 2000). In an open area, a bystander 1 m away will intercept no more than an average of about 0.05 W from such a transmission.

However, the entire point of the present work is not that some thermal or electrolytic effects are small, but rather that the <u>non</u>thermal effects of some EMR transmissions require much more investigation.

## *Dipole Moment Arguments*

Foster (2000) defines *E*-field torque by $|torque| = E\mu\cos\theta$, with $\mu$ a dipole moment and $\theta$ the angle between the field and the dipole. Then, his Eq. (20) represents the response time constant $\tau$ for dipole alignment as, $\tau = 4\pi\eta a^3/(kT)$, in which *a* is the dipole element radius (membrane half-width) and $\eta$ is a Stokes-law viscosity presumed to cause the interaction to be inelastic by thermal energy loss. Foster's Table II applies this formula to various biological bipolar structures, with the result that very large field strengths seem to be required for significant energy transfer to the structure. Foster reasons that only heat will be generated by such rotation because his model contains only a viscosity (dissipative) parameter.

Entirely ignoring the discussion above concerning time-domain polarization change, there are several other specific difficulties with this argument. The reasoning seems to be that if the time-constant $\tau$ above is too low, then the dipole moment must be small, so very little torque can be applied, and the mechanical interaction of the field with the membrane will be small. If the time-constant is too high, the dipole can not reorient and perhaps just scatters the EMR or is transparent to it.

EMR causes dipoles to rotate; the energy associated with the angular momentum transfer is converted from the energy of the incident photons. Dimensional analysis of the equation for $\tau$ above shows that the numerator may be seen as the product of a momentum and a distance -- which is to say, it represents a factor of an angular momentum. The distance in the numerator might be seen as a lever arm, and the momentum as that transferred at the point of application of a force. The energy associated with this momentum transfer is converted from the energy of the incident photons. For coherent EMR applied to a tissue boundary as in Fig. 5 above, there is no reason why the resulting mechanical wave in a population of such dipoles could not exert a pressure exceeding the radiation pressure (linear momentum) transferred directly to the dipole population by the incident microwave photons.

One difficulty in the dipole-moment approach of the Foster review is that it depends on individual dipoles in what is effectively a free, gaseous state.  It can not be applied accurately even to liquid pure water at room temperature, because of the inherent long-range structure of water.  The free parameter $\eta$ can be estimated *effectively* for water at an absorptive resonance point, but it can not be interpreted microscopically to represent the "viscosity" associated with a water molecule dipole in such a context.  Thus, it can not be cited reliably in attributing either thermal or nonthermal interaction to the dipole.

Also, Stoke's law strictly applies only to constant velocities, not to perpetually accelerated, reversing oscillations; so, it should not be assumed microscopically accurate in a single-molecule context.

Another difficulty is that when Foster's adaptation of this approach is applied to large, elongated elements such as DNA molecules, they are treated as though they were rigid bodies.

In the present author's opinion, the small value of $\tau$ as interpreted by Foster probably should be read as implying only that nonthermal energy was being ignored.  A better interpretation would seem to be that, with a small value of $\tau$, the membrane dipole can rotate freely and that therefore the membrane will have plenty of time to relax while receiving mechanical free energy.  This was pointed out in the section above on time-domain problems.

The argument from the time constant then depends on an EMR frequency far above that of a membrane resonance.  For example, Foster (2000, Table II) gives 20 GHz as the frequency corresponding to relaxation time $\tau$ of liquid water.  Therefore, EMR at, say, 5 GHz should have no difficulty rotating a water-sized dipole in a membrane by some considerable amount determined not by a constant viscosity, but by nonconstant elastic restoring forces applied to the dipole by nearby bonds in the membrane.  See Fig. (5) and Fig. (7) above.

# Nonthermal Causation

At this point, we turn to some evidence of specifically nonthermal biological interactions.  The studies described below are selected for relevance to the theme of the present work.  Because the present theme solely is a proper understanding of the fundamental physical interaction between microwaves and elements of tissue, many entirely valid biochemical and microbiological studies have been omitted; such may be found in the references of works cited, or in the reviews listed at the beginning of the present work.

To understand the variety of ways experiments are reported, it is useful to recall that for plane-wave EMR, power $P = E^2/Z$, in which $E$ is electric field strength and $Z$ is impedance.  In SI units, $P$ is in $\text{watts}/\text{m}^2$, $E$ is in volts/m, and the plane-wave impedance of vacuum or air is $Z = 120\pi = 377$ ohms.  Thus, a plane-wave EMR field of 100 V/m delivers $P = 26.5$ W/m$^2$ or 2.65 mW/cm$^2$.  To determine absorbed energy, the SAR (Specific Absorption Rate) of the tissue involved must be known in the appropriate frequency range; the result usually will be a spatiotemporal average power expressed in mW/kg of tissue.

## *Some Evidence*

Frey (1962) describes various auditory and tactile sensations including a feeling of "buffeting" of the head, and "pins and needles", when stimulating humans with audio-range pulse modulated RF between 0.4 and 3 GHz at average power density under 1 $\text{mW}/\text{cm}^2$.  He also reports headaches from microwave exposure (Frey, 1998).  For microwave hearing, as confirmed in Frey and Messenger (1973), the pulse peak power, not average power, determined the detection threshold.  The Frey and Messenger auditory response apparently depended upon electronic noise or high-order oscillator harmonics riding on the pulses.

Frey (1962) estimates that in an audio noise-free environment, a human could detect this kind of transmission at power levels of 3 $\mu\text{W}/\text{cm}^2$.  Frey notes that, assuming that the auditory system responds to EMR reduced already by a factor of ten by absorbance of the surrounding tissue, the auditory receiving-site input would be equivalent to about 300 $\text{nW}/\text{cm}^2$, a sensitivity within one order of magnitude of that of a 1960's vintage broadcast radio receiver.  This of course includes any possible macroscopic antenna-like functionality of the body parts as described above.  Ingalls (1967) writes that he was able to confirm Frey's thresholds approximately.

Directly demonstrating a nonthermal, not even biological, effect of EMR is the report by Sharp, *et al* (1974), who used GHz-range, pulsed radar to show that certain ordinary materials, such as crumpled aluminum foil or microwave-absorbing foam rubber, could emit sounds audible to nearby human observers.  The Sharp, *et al* pulses were around 10 $\mu$s wide, and their power at audible threshold was around 50 mW/cm$^2$ per pulse.  These values are below those of Frey and Messenger's (1973) or FF74's typical microwave hearing parameters, even though the Sharp, *et al* sounds were being heard through the air some distance from the listener's head.  Because aluminum metal reflects or scatters almost all the power of an incident pulse, it can not easily be argued that expansion because of absorbed heat accounted for the sounds under this condition in their study.  Various sizes of object were tried, down to a few mm on a side (see also Justesen, 1974); Sharp, *et al* therefore would seem to have eliminated acoustic geometrical resonance (*cf.* of the head) as responsible for the sounds.  These authors calculate that radiation pressure (linear momentum of the microwave photons) could explain their results.

Guy, *et al* (1975) reported having confirmed the Sharp, *et al* observations approximately; they studied the effect closely with microwave-absorbing coatings on plastic discs.  They concluded that acoustic waves in the discs showed higher pressures than those calculated from the momentum of the microwave photons applied.  This is hardly surprising, because photons, being massless, necessarily deliver the minimum possible momentum per unit energy.  So, any form of energy, thermal or not, converted by the discs from EMR, would permit of the possibility of pressures higher than the radiation pressure.  What is important is that only random momentum can accompany random energy (heat); so, all truly thermoacoustic materials must provide structure which randomizes the momentum delivered, whether linear, or angular, or both.

The *E*-field experiments such as those of Westerhoff, *et al* (1986) and Sommer and von Gierke (1964) provide important data at the low-frequency extreme of the EMR spectrum.  Slowly varying *E* fields are similar in effect to low-frequency EMR irradiation.  The *E* field provides a potential in which tissue will have time to remain relaxed in its lowest energy state, and in which current (charge carriers) will be flowing with differentially slow rate of change over almost all the time of a laboratory measurement.  EMR of very long wavelength has a similar action; but, as frequency approaches the microwave region, the wave nature of the EMR photons begins to dominate the effect.  As the gigahertz range is approached, we enter the diffractive region for humans.  Sommer and von Gierke found *E*-field hearing thresholds far above those of Frey (1962); however, they did not use EMR stimulation, so the spatial interactions with tissue would be expected to be less than in the Frey studies.  One should note that Ingalls (1967) writes that he could not produce auditory sensations with electrostatic fields alone.

Sanders and Joines (1984) used heat-stressed rats with and without application of continuous (unpulsed) 591 MHz EMR at less than 15 mW/cm$^2$ to the animals' heads.  Unpulsed presentation eliminated any possible thermal rate contribution.  They found that the EMR had no measurable effect on body temperature, but that it caused dramatic depletion of brain ATP and related neural energy stores, beyond that normally caused by increased metabolism during hyperthermia.  They concluded that microwaves can reduce brain metabolic rate.  Presumably this would imply impaired brain function, although rat behavior was not measured.

Microwaves have been found to cause cellular heat-stress responses far more easily than other kinds of stress -- including stress caused by heat (M. Blank,  in Sage &

Carpenter, 2012, Section 7).

Generalizing on the cellular commonality shared among insects and humans, Panagopoulos and Margaritis (2002) exposed fruit flies to GSM (~900 MHz) cell phone radiation at very low levels for just six minutes per day during the several days it takes such flies to hatch from newly-fertilized eggs.  The levels averaged between about $7\,\mu W/cm^2$ and under $3\,mW/cm^2$.  The exposed groups in their adult life showed a loss in reproductive activity varying between 15% and 60%, depending on estimated irradiance.  The authors conclude that GSM radiation may be capable of serious biological damage to human cells.  While the theory proposed by these authors is not compatible with that proposed in the present work, their data may be taken to validate an interaction consistent with a nonthermal theory.

Lu, *et al* (1999) applied spark-gap generated UWB microwave pulses to rats at too low an average level to cause thermal effects and found that the blood pressure dropped reliably; heart rate was not affected.  The peak fields were close to 100 kV/m; but, with 1 ns pulse width and only 500 to 1,000 pulses per second, the duty cycle was just 0.5 to 1.0 $\times 10^{-6}$.  So, the average exposure was less than 0.3 mW/cm$^2$, delivering only an estimated 100 mW/kg rat body mass, well below recommended safety limits based on thermal effects only.  The exposure was for one six-minute period; the pulse shape given in the paper shows frequency components above 5 GHz, corresponding to wavelengths of a few cm in air, easily in the diffractive range for a rat's body.  Lu, *et al* calculated that their microwave pulses were below microwave hearing threshold for the rat, based on the threshold power summation law ("Bunsen-Roscoe law") common to all sensory organs; the summation had been observed for microwave hearing by Frey and Messenger (1973), Guy, *et al* (1975), and others.

The Lu, *et al* drop in blood pressure persisted at least for several weeks, indicating a long-enduring or perhaps permanent effect.  The blood pressure effect, as well as Frey's (1962) report of "pins-and-needles", is entirely consistent with loss of smooth muscle or nerve function caused by standing patterns of microwaves.  Because migraine headache may be caused by dilation (reduced depolarization of muscle cells and thus loss of smooth-muscle tonus) of arteries in the scalp (Berkow, 1987, p. 1355), Frey's (1998) headaches and Lu, *et al*'s blood pressure drops may very well share a common cause.

Hocking and Westerman (2001) reported an incident in which the head and upper body of a worker inadvertently were exposed to ~850 MHz digital (pulsed) communication tower microwaves at average levels well below thermally-based safety limits.  The exposure lasted a couple of hours; it caused general malaise and measurable dysfunction of the central nervous system, as well as neurologically confirmed, long-lasting loss of tactile sensitivity of the facial skin, the loss being most marked for sensation mediated by "C" fibers.  The worker recovered gradually over a period of months.

More recently, Salford, *et al* (2003; see also Adey, 2003) have reported anatomical changes in rat brain caused by two-hour exposures to continuous 915 MHz microwave irradiation.  Apparent neuronal degeneration in localized brain regions was demonstrated at irradiances of around 1–10 mW/cm$^2$, far below any threshold of harm based on a thermal argument. Salford, *et al* observed an apparent breach of the rat blood-brain barrier and suggest that the neuronal damage may have resulted from it. These results have not yet been replicated under circumstances studying EMR diffraction by the rat's anatomy and the confining cage.

In a double-blind analysis of variance experiment, Zwamborn, *et al* (2003) found that 2.1 GHz EMR, simulating digital base-station transmissions (UMTS/3G protocol) at a maximum field strength of 1 V/m, significantly bothered certain persons either representing themselves as "electrosensitive" or not, but that other protocols (GSM at 900 or 1800 MHz) had no discernible effect. Several reaction-time tasks apparently were facilitated by RF irradiation. It seems reasonable to assume that facilitation of simple tasks would have been because of selective neural hyperpolarization of inhibitory processes, most of the brain having an inhibitive function. Zwamborn, *et al* conclude that their effects were nonthermal.

## *Some Implications*

By Fourier transform theory, temporally narrow pulses imply wide frequency bands; and, wide frequency bands require high-frequency components with sharply defined spatiotemporal features. So, for standing patterns, wide-band microwaves would be associated with relatively sharp spatial patterns on organ boundaries. Briefly sustained sharp spatial patterns would affect nerve and muscle in a way equivalent to sudden temporal changes such as action potentials, as pointed out in the discussion of Eq. (1) above.

If we assume that nonthermal interactions only cause changes in membrane polarization at cell boundaries, then we can explain *microwave hearing* as perhaps a combination of thermal (FF74) and nonthermal effects ultimately causing a response in the cochlea. We also can explain Frey's (1962) anecdotal report of tactile sensations, some reports of radar harassment, and even reports of therapeutic sexual stimulation (*e. g.*, Gorpinchenko, *et al*, 1995). We easily can explain the findings by Lu, *et al* (1999) and Hocking and Westerman (2001).

In a phenomenological review, Hyland (2000) has speculated that nonthermal interactions might depend strongly on individual differences, this being a reason that clear effects have not yet been reproducible or recognized more widely. This does not apply to *microwave hearing*, which has been universally and easily reproducible; no one in the modern peer-reviewed literature has denied that it could be caused reliably by pulsed microwave EMR.

Hyland's suggestion is supported by the present approach: Simple individual differences such as size of head, length of limb, or volume of stomach or lungs would be dominant under the assumption that spatial interactions with organ boundaries determined the nonthermal effects. In addition, reproducibility in past studies may have been inhibited by misunderstanding of the basic physics of the interaction of microwaves with tissue, a problem hopefully solved in the present work.

## Cancer

On a macroscopic scale, it is not obvious how microwaves could damage the self-repairing genetic material of body cells more severely than the nongenetic material, if their effect solely was because of diffraction at tissue boundaries.

Ionizing radiation, such as ultraviolet light or gamma radiation, has wavelengths below one micrometer, well below the microwave domain.  Each ionizing photon or charged particle damages anything with which it interacts.  This damage to cell material occurs at random and may be permanent, occasionally selectively damaging genetic material in a way putting the cell irreversibly into a cancerous state.

Microwaves span thousands of cells with a single wavelength, the $E$ fields rotating dipoles and setting up alternating electrical currents.  Presumably, a dipole rotation which broke, rather than merely stretched, the bonds holding the dipole would damage the protein or other molecule involved, as well as the membrane holding it.

As already mentioned, EMR parameters distinguishing mere hyperpolarization from damage have not been studied well, mainly because of thermal confusion; and, thus, they can not be described here.  The same or higher EMR irradiance damaging membranes presumably would cause destructive heating or local electrolysis.  In terms of tissue or organ boundaries, or diffraction, then, it would seem that a microwave current which damaged the genetic material also would damage everything around it, too, leaving little for the damaged genetic material to use for cancerous proliferation.

But, there are different inferences which are possible: On a microscopic scale, EMR might cause cells to become cancerous by direct action on dipole elements in protein molecules.  A protein may shift between metabolically active and inactive states merely by a small refolding.  The dipoles in any molecule as complex as a protein will be irregularly distributed.  So, inserting energy irregularly throughout such a protein might cause it to change its folding state -- its conformational state -- as has been shown at very low frequencies by Westerhoff, *et al* (1986).  The conformation depends heavily on hydrogen bonding, just as does the structure of water.  The barrier to a protein conformational change would be the stretch of a bond by a few bond-lengths (Bruscolini & Pelizzola, 2002).   Su and Heroux (2012) studied 60 Hz fields within cell-culture equipment and warned of this kind of interaction, citing several studies showing modification of cell development caused by such fields.  Because of the possibility of field-induced conformational changes, according to a British *Telegraph* newspaper report, an Italian court has ruled that prolonger daily use of a cell phone was the cause of a brain tumor in one individual (Alleyne, 2012).  The same report also mentions that the World Health Organization (WHO) has classified mobile telephones as Class B carcinogens.  Some fairly firm evidence that prolonged cell-phone use can increase the probability of leukemia in children is presented in various reports editted by Sage and Carpenter (2012) -- notably in their entire section 12.

However, why should not such subcellular changes also just as likely reduce the cancer rate (relative to diseases of other kinds)?  Yes, microwave EMR *might* activate certain enzymic changes possibly increasing the cancer rate; however, more specific hypotheses about oncogenesis should be formulated before such changes were to be linked meaningfully to EMR exposure.

## Vehicle Safety and Cell Phones

Rothman (2000) suggests that cell phones should be elevating the death rate from auto accidents far more than the rate from cancer.  In view of the effect of some pulsed microwaves on blood pressure (Lu, *et al*, 1999), one should hope for caution in introducing wideband cellphones into automobiles: They might trigger fainting or some other brief, unnoticed loss of consciousness at the wheel.  Many more recent reports of increased auto accident rate because of cell phone use simply depend upon distraction; however, one factor in this distraction would be, of course, effectiveness of the brain function while driving.  Also, even a brief reduction in blood pressure to the brain may cause mild stroke (transient ischemic attack) in a vulnerable individual (Berkow, 1987, p. 1382). Accidents from the EMR of wideband cell phones, so far as known to the author, can not currently be distinguished from accidents of inattention or sudden, presumably EMR-unrelated, illness.

The exact mechanism of the blood pressure and headache effects described above requires elucidation: A handheld device emitting narrow pulses at 10 kV/m near the antenna tip might deliver a localized field intensity equal to those of Lu, *et al*'s experiment, if held against a (diffractive) human head.

## Tinnitus

Microwave hearing resembles tinnitus, a symptom of cochlear damage.  Tinnitus is a sensation of a hissing, buzzing, or sizzling sound in the head (Meikle, *et al*, 2000); it may be intermittent and triggered by a loud noise.  Inner-ear electrical abnormalities may accompany tinnitus (Mayo Clinic, 2001).  The sometimes-identifiable pathology has been reviewed by Lockwood, et al (2002).

It seems certain that some cases of tinnitus, especially if intermittent, actually must be reports of microwave hearing.  The present author is not aware of any documented, diagnosed case of tinnitus later found to be microwave hearing.  However, a systematic field investigation at several locations in Germany, including some blinded trials, has reported that a variety of tinnitus-like symptoms were mitigated by shielding the head with aluminum foil but not with paper (Mosgoeller and Kundi, 2000).

Keeping in mind the possibly harmful side-effects, perhaps a study of microwave hearing in tinnitus patients would reveal special vulnerabilities in such persons.  Such a study might lead to insight permitting amelioration of the tinnitus, or possibly to new diagnostic techniques.

# Summary

We have indicated the importance of being able to investigate meaningfully, and thus resolve, claims of misuse of EMR in ways harmful to humans. We enumerated several of the current political problems hinging on the success of such investigations.

We then stepped back and examined the basics of EMR biological interaction, reviewing how EMR can cause temperature to rise in tissue. We pointed out that there must be nonthermal interactions at a cellular level at all tissue or organ boundaries. We also pointed out a second, diffractive, antenna-like interaction on the scale of the gross human anatomy. To lay groundwork for subsequent criticism, we explained some peculiarities of the structure of liquid water.

We presented the FF74 hypothesis, which describes microwave hearing purely as a thermoacoustic effect. We showed that, based on vibrations of the human head, this hypothesis could account for microwave hearing only slightly above absolute detection threshold. We showed that a study by Gournay, cited by FF74, did not independently support the hypothesis.

We pointed out that the crucial phase reversal at 4 C in the FF74 experiment did not necessarily demonstrate that a temperature pulse was occurring. We showed that a structural explanation of the phase reversal was possible and could be tested empirically. We then presented a complementary nonacoustic hypothesis of microwave hearing which assumed direct EMR interaction with the cochlea and which better fit the data than did the FF74 hypothesis.

Generalizing our criticism, we then addressed several basic misconceptions in the literature of bioeffects of microwaves. These included the use of a spherical model of a living cell, ignorance of time-domain patterns on tissue boundaries, invalid invocation of thermal noise arguments, and incomplete understanding of the meaning of an electric dipole on a boundary as opposed to one in the bulk of a fluid cavity.

Having come to the end of the criticism, we next listed several reliable experimental results demonstrating clear biological effects not explainable in terms of a thermal or thermal rate mechanism.

Finally, we elaborated a few implications of the nonthermal mechanisms presented, including possible effects related to cancer, to vehicle safety, and to the microwave-hearing-like auditory symptom, tinnitus.

# Conclusion

Bothersome or unhealthful effects of EMR within thermal safety limits have been demonstrated empirically and are supported theoretically.  The microwave spectrum in certain bands, or under certain conditions, probably should be treated more like a pharmacist's shelf than the temperature knob on an oven.  Decisions as to safety should be steered by persons more knowledgeable in medicine than has been the case in the past.

A particular worry, in light of the political controversy at the start of the present work, is that the widespread miscomprehension of biological effects of EMR implies that misuse of microwave devices can not be identified reliably by examination of a purported victim. Means do not exist, based on forensic evidence, for prosecuting malicious misuse causing distress, pain, injury, or death.

# Acknowledgements

The author thanks W. R. Adey and A. H. Frey for reading an early draft and offering suggestions.  Thanks especially are due to Dr. Bruce Hocking, who offered relevant case studies as well as advice considerably improving the exposition.

Some of the work was done while the author was enrolled in the *Stanford University School of Continuing Studies*, and much of the research cited was because of access to the Stanford University libraries.